\documentclass[preprint,amssymb,amsfonts,amsmath,superscriptaddress]{revtex4-2}
\usepackage[margin = 0.88in]{geometry}  
\usepackage{lineno}
\usepackage{xcolor}
\usepackage{enumitem}
\usepackage{dcolumn}
\usepackage{bm}
\usepackage[printwatermark]{xwatermark}
\usepackage[bookmarks=true, bookmarksopen = true, pdfstartview={FitH}, colorlinks = true, citecolor = blue, linkcolor = blue, urlcolor = blue]{hyperref}
\usepackage{dcolumn}

\DeclareFontFamily{U}{euc}{}
\DeclareFontShape{U}{euc}{m}{n}{<-6>eurm5<6-8>eurm7<8->eurm10}{}%
\DeclareSymbolFont{AMSc}{U}{euc}{m}{n} 
\DeclareMathSymbol{\umu}{\mathord}{AMSc}{"16}

\begin{document}

\def\MEAM{\small Department of Mechanical Engineering and Applied Mechanics, University of Pennsylvania, Philadelphia, PA 19104, USA}
\def\EES{\small Department of Earth and Environmental Science, University of Pennsylvania, Philadelphia, PA 19104, USA}

\title{Understanding the rheology of kaolinite clay suspensions using Bayesian inference}

\author{Ranjiangshang Ran}
\affiliation{\MEAM}
\author{Shravan Pradeep}
\affiliation{\EES}
\author{S\'ebastien Kosgodagan Acharige}
\affiliation{\MEAM}
\affiliation{\EES}
\author{Brendan C. Blackwell}
\affiliation{\MEAM}
\author{Christoph Kammer}
\affiliation{\MEAM}
\author{Douglas J. Jerolmack}
\affiliation{\EES}
\author{Paulo E. Arratia}
\thanks{Corresponding author: parratia@seas.upenn.edu}
\affiliation{\MEAM}
\date{\today}

\begin{abstract}
Mud is a suspension of fine-grained particles (sand, silt, and clay) in water. The interaction of clay minerals in mud gives rise to complex rheological behaviors, such as yield stress, thixotropy and viscoelasticity. Here, we experimentally examine the flow behaviors of kaolinite clay suspensions, a model mud, using steady shear rheometry. The flow curves exhibit both yield stress and rheological hysteresis behaviors for various kaolinite volume fractions ($\phi_k$). Further understanding of these behaviors requires fitting to existing constitutive models, which is challenging due to numerous fitting parameters. To this end, we employ a Bayesian inference method, Markov chain Monte Carlo (MCMC), to fit the experimental flow curves to a microstructural viscoelastic model. The method allows us to estimate the rheological properties of the clay suspensions, such as viscosity, yield stress, and relaxation time scales. The comparison of the inherent relaxation time scales suggests that kaolinite clay suspensions are strongly viscoelastic and weakly thixotropic at relatively low $\phi_k$, while being almost inelastic and purely thixotropic at high $\phi_k$. Overall, our results provide a framework for predictive model fitting to elucidate the rheological behaviors of natural materials and other structured fluids.

\end{abstract}
\maketitle

\newpage

\section{Introduction}
Mud --- a suspension of fine-grained particles (such as sand, silt, and clay) and water --- forms the foundation of Earth's surface and plays important roles in many natural (e.g., mudslides \cite{Hungr_Landslides_2014}, soil erosion \cite{Pimentel_Soil_2006}, river morphology \cite{dunne2020sets,phillips2022threshold}) and industrial processes (e.g., civil construction \cite{Barnes_Civil_2000}, oil drilling \cite{VanDyke_drilling_1998}). At low stresses, mud exhibits solid-like behavior, which gives the material its load-bearing capacity. Above a certain threshold stress, however, these suspensions can experience sudden mechanical failure and start to flow like a liquid, as seen during the rapid and catastrophic mudslides \cite{Jerolmack_NRP_2019}. This threshold at which the transition from solid- to fluid-like behaviors occurs is referred to as the ``yield stress'' \cite{Coussot_JOR_2002,Cruz_PRE_2002,Bonn_RMP_2017,Frigaard_COCIS_2019}. Ultimately, to understand yielding behaviors in mud-like materials, we need models that can predict the yield stress of mud suspensions based on its constituents and bulk rheology. If successful, such rheological models could help improve failure predictions for industrial processes and prevent hazardous natural events.

The rheology of mud is dominated by the attractive nature of clay minerals, since these charged plate-shaped mud constituents are observed to aggregate and form gel-like microstructure \cite{Coussot_PRL_1995,Zukoski_JOR_1999,Brown_JOR_2000,Zbik_Kaolin_2008,Gupta_JCIS_Struc_2011,Phan_Thien_Langmuir_2015,Neelakantan_Mineral_2018}. This gives rise to ``thixotropy'' \cite{Mewis_JNNFM_1979,Barnes_JNNFM_1997}, a rheological behavior characterized by time-dependent viscosity resulting from the evolution of the microstructure associated with flow history \cite{Mewis_ACIS_2009}. Thixotropic behaviors often involve two processes: an increase in viscosity due to the buildup of microstructure as the fluid is at rest (or ``aging''), and a decrease in viscosity due to the breakdown of microstructure as the fluid is sheared (or ``rejuvenation'') \cite{Larson_JOR_2015,Larson_JOR_2019}. Rheological models of thixotropy often use a structural parameter to relate the fluid microstructure to its macroscopic properties (e.g., viscosity), along with a kinetic equation to describe the evolution of the structural parameter \cite{Moore_1959,Stickel_Model_JOR_2006,Grillet_JOR_2009,Mewis_JNNFM_2006,DeSouza_JNNFM_2009,DeSouza_SoftMatt_2011,DeSouza_JNNFM_2012,DeSouza_JNNFM_2013,Blackwell_JNNFM_2014,Blackwell_JNNFM_2016,Wagner_JOR_2016,Larson_JOR_2016,Larson_JOR_2018,Joshi_JOR_2021,Joshi_JOR_2022}. A similar structural parameter and accompanied kinetic equation are also used to describe the rheological behaviors of ``soft glassy'' materials that are usually not referred to as thixotropic \cite{SGR_PRL_1997,SGR_PRE_1998,Bocquet_PRL_2009,Bocquet_SoftMatt_2011,Jeyaseelan_Model_JNNFM_2008,Fielding_PRL_2011,Fielding_SM_2017,Fielding_PRL_2013}. While the structural parameter is often taken to be a single scalar quantity \cite{Grillet_JOR_2009,Mewis_JNNFM_2006,DeSouza_JNNFM_2009,DeSouza_SoftMatt_2011,DeSouza_JNNFM_2012,DeSouza_JNNFM_2013,Blackwell_JNNFM_2014,Blackwell_JNNFM_2016,Wagner_JOR_2016,Joshi_JOR_2021,Joshi_JOR_2022}, it can also be represented as tensors \cite{Stickel_Model_JOR_2006,Jeyaseelan_Model_JNNFM_2008,Fielding_PRL_2013}, scalar fields \cite{Fielding_PRL_2011,Fielding_SM_2017}, or a combination of multiple scalar quantities \cite{Larson_JOR_2016,Larson_JOR_2018}. Despite its clear physical meaning, the estimation of the structural parameter from experimental measurements is challenging.

Besides yield stress and thixotropy, mud and other clay-based suspensions can also exhibit viscoelasticity \cite{Bossard_JOR_2003,Bossard_JOR_2007,Joshi_PRSA_2008,Joshi_RA_2018}, since clay particles can form gel-like microstructure that behaves like weak elastic solids \cite{Mewis_ACIS_2009,Zukoski_JOR_1997,Kaolin_PoF_2021}. Viscoelastic behaviors are characterized by the relaxation of elastic stress over an observed time scale \cite{Larson_rheology_1999}. The time scales associated with stress relaxation and microstructural evolution are usually different and independent, but it is often difficult to distinguish them \cite{Mewis_ACIS_2009,Larson_JOR_2015,Larson_JOR_2019,Joshi_JOR_2021}. Therefore, to capture the rheological behaviors of mud and other structured fluids, such as food products \cite{Augusto_Food_2012,Glicerina_Chocolate_2015,Glicerina_Chocolate_2016,Potato_CP_2016} and crude oils \cite{Mckinley_SoftMatt_2014,McKinley_JOR_2017}, constitutive models that incorporate both thixotropy and viscoelasticity are necessary \cite{Mewis_JNNFM_2006,DeSouza_JNNFM_2009,DeSouza_SoftMatt_2011,DeSouza_JNNFM_2012,DeSouza_JNNFM_2013,Blackwell_JNNFM_2014,Blackwell_JNNFM_2016,Wagner_JOR_2016,Larson_JOR_2018,Joshi_JOR_2021,Joshi_JOR_2022,SGR_PRL_1997,SGR_PRE_1998,Bocquet_PRL_2009,Bocquet_SoftMatt_2011,Jeyaseelan_Model_JNNFM_2008,Fielding_PRL_2011,Fielding_SM_2017,Fielding_PRL_2013,Mckinley_SoftMatt_2014,McKinley_JOR_2017}. Such models commonly contain Maxwell- or Jeffrey-type viscoelastic elements whose viscosity and elasticity depend on the structural parameter. An exception is when the viscoelastic time scale (i.e., stress relaxation) is much shorter than that of thixotropic time scale (i.e., microstructural evolution), in which case the material is called ``ideally'' thixotropic and the effects of viscoelasticity can be neglected \cite{Larson_JOR_2016,Larson_JOR_2015,Larson_JOR_2019}.

The interplay of thixotropy and viscoelasticity also leads to hysteresis behavior \cite{Divoux_PRL_2013,Puisto_PRE_2015,Fielding_SM_2017,Fielding_JOR_2020,Mckinley_PRL_2019}. A common procedure to identify such behavior is by sweeping the shear rate (or shear stress) downward and upward over a prescribed range \cite{Divoux_PRL_2013,Fielding_SM_2017}, which produces hysteresis loops in stress versus shear rate flow curves. This phenomenon is termed rheological hysteresis and has been widely reported in soft glassy materials such as clay, cement, and mud suspensions \cite{Beris_JNNFM_2002,Divoux_PRL_2013,Puisto_PRE_2015,Fielding_JOR_2020,Mckinley_PRL_2019,Divoux_Soft_Matt_2011,Fourmentin_Acta_2015, perret_1996_thixotropic,jeong_2015_thixotropic, labanda_2005_influence}. One way to quantify rheological hysteresis is by calculating the area enclosed by the hysteresis loops, i.e., hysteresis area \cite{Divoux_Soft_Matt_2011,Divoux_PRL_2013,Fielding_SM_2017,Mckinley_PRL_2019,Jamali_JoR_2022,Joshi_ArXiv_2022}. Since the rheological hysteresis depends on the magnitude and time step of the applied shear, the non-monotonic dependence of the hysteresis area as a function of time step can be used to infer the material’s intrinsic relaxation time scale \cite{Divoux_PRL_2013,Fielding_SM_2017,Mckinley_PRL_2019,Jamali_JoR_2022}. While such time scale has been associated with thixotropy, recent numerical simulations also observe the non-monotonic dependence of hysteresis area in purely viscoelastic materials \cite{Joshi_ArXiv_2022}. Theoretically, the intrinsic time scales in materials that exhibit rheological hysteresis was studied using a fluidity model and the soft glassy rheology (SGR) model, which reproduce similar hysteresis behavior observed in the experimental systems \cite{Puisto_PRE_2015,Fielding_SM_2017}. Alternatively, large amplitude oscillatory shear (LAOS) method is also reported in the literature to characterize the rheological hysteresis \cite{Ewoldt_JOR_2008,Ewoldt_JOR_2013,Blackwell_JNNFM_2014,Blackwell_JNNFM_2016,Wagner_JOR_2016,SARogers_PRL_2019,SARogers_PNAS_2020}. Despite the advances in rheological models and measurements, it remains challenging to fit constitutive models to the experimental flow curves obtained from steady shear or LAOS, due to the model complexity and numerous fitting parameters. Therefore, common model fitting approaches often apply statistical methods such as Bayesian inference \cite{Ewoldt_JOR_2015} and neural networks \cite{Jamali_JOR_2021,Jamali_SciRep_2021,Jamali_SoftMatt_2022,Jamali_PNAS_2022}.

In this study, we experimentally examine the flow behaviors of kaolinite clay suspensions --- a model mud --- using shear rate sweep tests. The flow curves exhibit both yield stress and hysteresis behaviors for different kaolinite volume fractions ($\phi_k$). In an effort to understand these results, we fit the experimental flow curves to a constitutive model that is both viscoelastic and thixotropic, using Markov chain Monte Carlo (MCMC), a Bayesian inference method. The model fitting results provide estimates of the rheological properties of the clay suspensions, such as viscosity, yield stress, and inherent relaxation time scales. A comparison of the relaxation time scales of elastic stress and microstructure suggests that kaolinite clay suspensions are strongly viscoelastic and weakly thixotropic at relatively low $\phi_k$, while they are almost ideally thixotropic and inelastic at high $\phi_k$. Our results also reveal that kaolinite clay suspensions possess multiple time scales of stress and microstructural relaxation. Overall, our results provide insights into the development of the rheological models of natural materials such as mud and clay, as well as other structured fluids.

\section{Materials and Methods}

\subsection{Kaolinite clay suspensions}
Kaolinite clay suspensions are prepared by dispersing natural kaolinite particles (03584 Sigma-Aldrich) into deionized water at four different volume fractions: $\phi=$ 0.05, 0.12, 0.19, and 0.26. Kaolinite is a layered aluminum silicate ($\mathrm{Al}_2\mathrm{O}_3\cdot2\mathrm{SiO}_2\cdot2\mathrm{H}_2\mathrm{O}$), which consists of stacks of individual plate-shaped particles (particle density, $\rho_p=2.61~\mathrm{g/cm}^3$) joined together by hydrogen bonds. Kaolinite particles have a modal particle size, $d\approx10~\umu$m \cite{Seiphoori_GRL_2021,Phan_Thien_Langmuir_2015}, and an aspect ratio $d/h\approx10$, where $h$ is the average thickness of the particles. The zeta potential of kaolinite particles at $\mathrm{pH}=7.0$ is estimated to be $-30$~mV \cite{Seiphoori_GRL_2021}, which indicates that kaolinite particles are weakly attractive and tend to aggregate in water.

\subsection{Shear rate sweep test}
\begin{figure}[t!]
\centering
\includegraphics[width = 6.69 in]{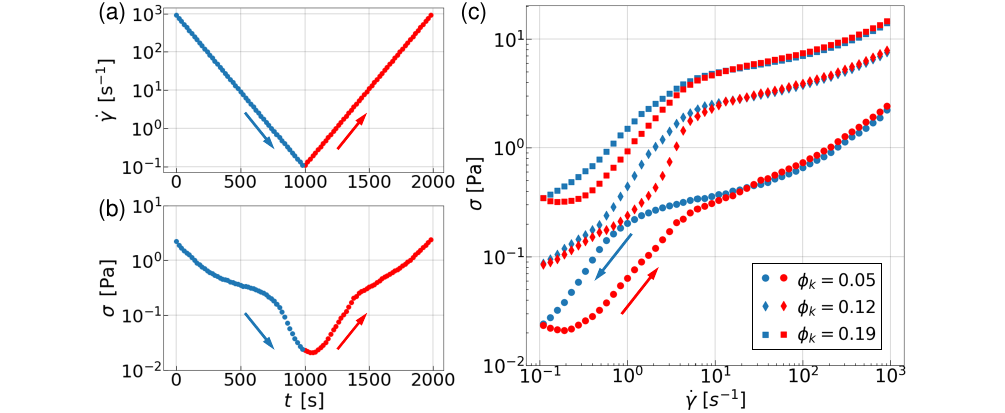}
\vspace{-3mm}
\caption{\small (a) The prescribed shear rate ($\dot{\gamma}$) as a function of time in the shear rate sweep test: a downsweep (blue) from $\dot{\gamma}_\mathrm{max}=10^3~\mathrm{s}^{-1}$ to $\dot{\gamma}_\mathrm{min}=10^{-1}~\mathrm{s}^{-1}$ in 50 logarithmic steps of duration $\Delta t=10~$s is followed by an upsweep (red) from $\dot{\gamma}_\mathrm{min}$ back to $\dot{\gamma}_\mathrm{max}$ with the same steps in reverse order. (b) Experimental data for shear stress ($\sigma$) as a function of time $t$, for kaolinite suspension of $\phi_k=0.05$ and $\Delta t=10~$s, showing an asymmetry in time. (c) Flow curves $\sigma$ \textit{vs.} $\dot{\gamma}$ for kaolinite suspensions of different volume fractions: $\phi_k=0.05$, 0.12, 0.19, and $\Delta t=10~$s, showing rheological hysteresis behaviors.}
\label{fig1}
\end{figure}

A stress-controlled cone-plate rheometer (Bohlin, Malvern) is used to perform shear rate sweep tests that are similar to the protocols developed by Divoux \textit{et al.} \cite{Divoux_PRL_2013}. First, the sample is pre-sheared at a high shear rate ($\dot{\gamma}_\mathrm{max}$) for 180~s to eliminate previous flow history and generate a reproducible initial microstructure. Next, two consecutive shear rate sweeps are performed: a downward sweep from the highest shear rate ($\dot{\gamma}_\mathrm{max}=10^3~\mathrm{s}^{-1}$) to the lowest shear rate ($\dot{\gamma}_\mathrm{min}=10^{-1}~\mathrm{s}^{-1}$) through $N=50$ logarithmically spaced steps of duration $\Delta t$ each, followed by an upward sweep from $\dot{\gamma}_\mathrm{min}$ back to $\dot{\gamma}_\mathrm{max}$ with the same $N$ logarithmic steps in a reverse order. In this manuscript, we adopt the terms ``upsweep'' for the upward sweep in shear rates ($ \dot{\gamma}_\mathrm{min} \rightarrow \dot{\gamma}_\mathrm{max}$) and ``downsweep'' for the downward sweep in shear rates ($ \dot{\gamma}_\mathrm{max} \rightarrow \dot{\gamma}_\mathrm{min}$), as used in previous literature \cite{Fielding_SM_2017}. Two different sweep time $\Delta t$ are used: $\Delta t=10$~s and 20~s, leading to total experimental duration of $2N\Delta t=2000$~s and 4000~s, respectively. Figure \ref{fig1}(a) shows the prescribed shear rate as a function of time for $\Delta t=10~$s. The room temperature is maintained at 21$~{}^{\circ}$C during all the tests.

Figure \ref{fig1}(b) shows the measured stress in a shear rate sweep test, for kaolinite suspension of volume fraction $\phi = 0.05$ and $\Delta t=10~$s. We find that the suspension stress is asymmetric in time even though the applied shear rate is symmetric in time \ref{fig1}(a); this asymmetry leads to rheological hysteresis in the $\sigma$-$\dot{\gamma}$ flow curves. Figure \ref{fig1}(c) shows examples of the rheological hysteresis behaviors for kaolinite suspensions of different volume fractions: $\phi= $0.05,\ 0.12, and 0.19 with $\Delta t=10~$s. At low $\dot{\gamma}$, the suspension stress in upsweep (red) lies below the downsweep (blue) but beyond a critical shear rate ($\dot{\gamma_c}$) the flow curves reverse their behavior. 

\subsection{Rheological model}\label{sec. model}

\begin{figure}[t!]
\centering
\includegraphics[width=3.37in]{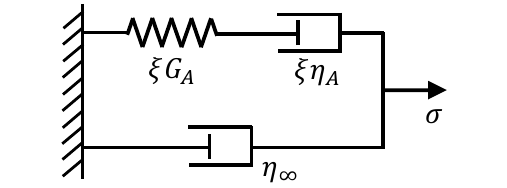}
\vspace{-3mm}
\caption{\small A mechanical analog of the canonical three-element viscoelastic thixotropic model.}
\label{fig2}
\end{figure}

A viscoelastic thixotropic model \cite{Blackwell_JNNFM_2014} is used here to capture the clay suspensions' rheological behavior described above. The model is represented by a simple three-element Jeffreys canonical form in Fig. \ref{fig2}, where a constant infinite shear viscosity ($\eta_\infty$) is modeled in parallel with a viscoelastic Maxwell component that is dependent on a scalar structural parameter, $\xi$. The viscous and elastic parts of the Maxwell component are assumed to be linearly dependent on the structural parameter as $\xi\eta_A$ and $\xi G_A$, respectively, where $\eta_A$ is the structural viscosity and $G_A$ is the elastic modulus. The time evolution of the shear stress $\sigma$ under an applied shear rate $\dot{\gamma}$ in a Jeffreys-type fluid can be described as:
\begin{equation}\label{Jeffreys model}
    \sigma+\lambda\frac{d\sigma}{dt}=\left(\eta_\infty+\xi\eta_A\right)\left(\dot{\gamma}+\lambda_r\frac{d\dot{\gamma}}{dt}\right),
\end{equation}
where the fluid has a total viscosity $\eta=\eta_\infty+\xi\eta_A$, a single relaxation time $\lambda=\eta_A/G_A$ independent of microstructure, and a microstructural-dependent retardation time $\lambda_r=\lambda\eta_\infty/(\eta_\infty+\xi\eta_A)$. Equation (\ref{Jeffreys model}) can be further reduced to the following form \cite{Blackwell_JNNFM_2014,Blackwell_JNNFM_2016}:
\begin{equation}\label{eqn Blackwell}
    \frac{\sigma}{\eta_A}+\frac{1}{G_A}\frac{d\sigma}{dt}=\left(\xi+\frac{\eta_0}{\eta_A}\right)\dot{\gamma}+\frac{\eta_\infty}{G_A}\frac{d\dot{\gamma}}{dt}.
\end{equation}
We assume homogeneous flow conditions in our model, and thus Equations \eqref{Jeffreys model} and \eqref{eqn Blackwell} are given in the scalar form.

The time-evolution of the suspension microstructure is described using a kinetic equation of the form \cite{Moore_1959}:
\begin{equation}\label{eqn Moore}
    \frac{d\xi}{dt}=k_A(1-\xi)-k_R\left\vert\dot{\gamma}\right\vert\xi,
\end{equation}
where $\xi$ is a dimensionless structural parameter that ranges between 0 and 1 representing the unstructured and fully structured states, respectively. Equation \eqref{eqn Moore} describes two different processes associated with the evolution of microstructure, namely ``aging'' and ``rejuvenation'', represented by the kinetic rate constants, $k_A$ and $k_R$, respectively. Aging is a spontaneous buildup of microstructure when the fluid is undisturbed; and rejuvenation refers to a flow-induced breakdown of microstructure when the fluid is under shear \cite{Mewis_ACIS_2009}. To minimize the number of fitting parameters, we use the combination of Equations (\ref{eqn Blackwell}) and (\ref{eqn Moore}) to provide a minimal model of viscoelastic thixotropy. In this model, there is a total of 5 fitting parameters: infinite-shear viscosity ($\eta_\infty$), structural viscosity ($\eta_A$), elastic modulus ($G_A$), aging rate constant ($k_A$), and rejuvenation rate constant ($k_R$). Despite the simplicity of this model, we still need a statistical inference method to fit it to the experimental flow curves, as discussed in Sec. \ref{sec. MCMC}.

\subsection{Markov chain Monte Carlo (MCMC) method} \label{sec. MCMC}

Due to the relatively large number of fitting parameters, we need a statistical inference method to fit the proposed rheological model to our experimental data. To this end, we use the Markov chain Monte Carlo (MCMC) method, which is a class of algorithms that can generate a sequence of random samples from a probability distribution where direct sampling is difficult. To setup the problem, we consider a model $\mathcal{M}$ determined by a set of parameters $\vec{\theta}$ that can make predictions on data $\vec{\mathcal{D}}$. Here, $\mathcal{M}$ is the five-parameter rheological model proposed in the previous section with parameters $\vec{\theta}=[\eta_\infty,\eta_A,G_A,k_A,k_R]$, and the data $\vec{\mathcal{D}}=\{\sigma,\dot{\gamma}\}$l; where the latter contains the experimentally measured shear stress and shear rate. The parameters $\vec{\theta}$ have a probability density $\mathbb{P}(\vec{\theta})$ satisfying some prior distributions, or initial guess. The distribution of interest is the posterior distribution of the density $\mathbb{P}(\vec{\theta}\vert\vec{\mathcal{D}})$, or the probability that the parameters $\vec{\theta}$ can provide the best model $\mathcal{M}$ given the data $\vec{\mathcal{D}}$. We can compute this probability from Bayes' theorem:
\begin{equation}\label{eqn. posterior}
    \mathbb{P}(\vec{\theta}\vert\vec{\mathcal{D}}) = \frac{\mathbb{P}(\vec{\mathcal{D}}\vert\vec{\theta})\mathbb{P}(\vec{\theta})}{\mathbb{P}(\vec{\mathcal{D}})},
\end{equation}
where $\mathbb{P}(\vec{\mathcal{D}})=\int\mathbb{P}(\vec{\mathcal{D}}\vert\vec{\theta})\mathbb{P}(\vec{\theta})\,d\vec{\theta}$ is a normalizing constant \cite{MCMC_PNAS_2003}. The parameters that can best fit the experimental data are the ones maximizing Equation \eqref{eqn. posterior}, which is given as:
\begin{equation}
    \vec{\theta}_{\mathrm{best}}=\underset{\vec{\theta}}{\mathrm{argmax}}\left\{ \mathbb{P}(\vec{\mathcal{D}}\vert\vec{\theta})\mathbb{P}(\vec{\theta})\right\},
\end{equation}
and also known as the maximum likelihood estimation (MLE).

For the five-parameter model used here, an explicit formula of the posterior probability $\mathbb{P}(\vec{\theta}\vert\vec{\mathcal{D}})$ is difficult to obtain, and direct sampling from $\mathbb{P}(\vec{\theta}\vert\vec{\mathcal{D}})$ is computationally prohibitive. Hence, we resort to stochastic sampling methods (MCMC) to generate observations from the posterior distribution \cite{Ewoldt_JOR_2015}. Specifically, we use the Metropolis-Hastings (M-H) algorithm that was developed by Metropolis \textit{et al.} \cite{Metropolis_1953} and later generalized by Hastings \cite{Hastings_1970}. A standard procedure of the M-H algorithm to generate a Markov chain consisting of the following steps:
\begin{itemize}[leftmargin=0pt]
\item[] Step 0. Propose an initial guess $\vec{\theta}=\vec{\theta_0}$.
\item[] Step 1. Denote the current state as $\vec{\theta}$; propose a new state $\vec{\theta^\prime}$ according to a transition kernel $q(\vec{\theta}\to\vec{\theta^\prime})$.
\item[] Step 2. Calculate the acceptance ratio transitioning from $\vec{\theta}$ to $\vec{\theta^\prime}$:
\begin{equation}
    \alpha=\min\left({1,\frac{\mathbb{P}(\vec{\mathcal{D}}\vert\vec{\theta^\prime})\mathbb{P}(\vec{\theta^\prime})}{\mathbb{P}(\vec{\mathcal{D}}\vert\vec{\theta})\mathbb{P}(\vec{\theta})}}\right).
\end{equation}
\item[] Step 3. Draw an arbitrary number $u$ from a uniform distribution $\mathcal{U}[0,1]$. If $u<\alpha$, accept the new state $\vec{\theta}=\vec{\theta^\prime}$; otherwise reject $\vec{\theta^\prime}$ and keep $\vec{\theta}=\vec{\theta}$. Return to Step 1.
\end{itemize}
Here, the transition kernel $q(\vec{\theta}\to\vec{\theta^\prime})$ is chosen to be a random walk:
\begin{equation}
    q(\vec{\theta}\to\vec{\theta^\prime}): \vec{\theta^\prime} = \vec{\theta}+\mathcal{A}\vec{\epsilon},
\end{equation}
where $\vec{\epsilon}\sim\mathcal{N}(0,\epsilon_{\mathrm{rw}}^2\bm{I})$ is a normally distributed dimensionless perturbation vector, with $\epsilon_{\mathrm{rw}}$ being a measure of step size of the random walk and $\bm{I}$ being the identity matrix; $\mathcal{A}$ is a diagonal matrix adapting the step size in different dimensions \cite{Adaptive_MCMC_2001,Adaptive_MCMC_2008}, since all five parameters of $\vec{\theta}$ have different orders of magnitude and units. 

The prior distribution of $\vec{\theta}$ is assumed to be normal: $\vec{\theta}\sim\mathcal{N}(\vec{\theta_0},\epsilon_{\theta}^2\mathcal{A})$, where $\epsilon_{\theta}$ is a measure of uncertainty in the prior distribution, and $\mathcal{A}$ is diagonal so that the components of $\vec{\theta}$ are independent. The posterior distribution of $\vec{\theta}$ relies on the definition of $\mathbb{P}(\vec{\mathcal{D}}\vert\vec{\theta})$. We denote the experimental data as $\vec{\mathcal{D}}$, and the data generated by model $\mathcal{M}$ with parameters $\vec{\theta}$ as $\vec{\mathcal{D}^\prime}$. We can then calculate the ``distance'' between $\vec{\mathcal{D}}$ and $\vec{\mathcal{D}^\prime}$ by a metric $\rho$, defined as:
\begin{equation}\label{eqn. loss func}
    \rho(\vec{\mathcal{D}},\vec{\mathcal{D}^\prime}) = \left\Vert\log(\sigma)-\log(\sigma^\prime)\right\Vert,
\end{equation}
where $\left\Vert\cdot\right\Vert$ denotes the $L^2$-norm, and $\sigma$ and $\sigma^\prime$ are the experimental and model predicted shear stresses, respectively. Here, we use the logarithmic stress $\log(\sigma)$ to put an equal weight to the stress data of low shear rates, which is order of magnitude smaller than that of high shear rates. Note that the structural parameter $\xi$ is not included in Equation \eqref{eqn. loss func} despite being predicted by the model $\mathcal{M}$, because $\xi$ is unobservable in the experiments. Alternative methods quantifying the prediction of unobserved quantities can be found in \cite{Unobserved_2015}. Ideally, the metric $\rho(\vec{\mathcal{D}},\vec{\mathcal{D}^\prime})$ is zero for a ``perfect'' model. We assume the metric is normally distributed with a zero mean: $\rho(\vec{\mathcal{D}},\vec{\mathcal{D}^\prime})\sim\mathcal{N}(0,\epsilon_\rho^2)$, therefore we have the conditional probability:
\begin{equation}
    \mathbb{P}(\vec{\mathcal{D}}\vert\vec{\theta})=\left(1/\sqrt{2\pi}\epsilon_\rho\right)\exp\left(-\rho^2(\vec{\mathcal{D}},\vec{\mathcal{D}^\prime})/2\epsilon_\rho^2\right).
\end{equation}

\begin{figure}[t!]
\centering
\includegraphics[width=3.37in]{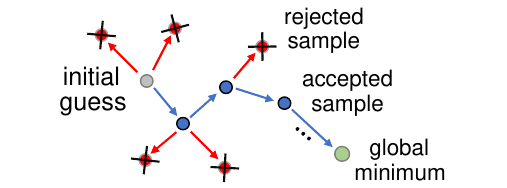}
\vspace{-3mm}
\caption{\small A graphical representation of the MCMC method, M-H algorithm.}
\label{fig.MCMC}
\end{figure}

\begin{figure}[t!]
\centering
\includegraphics[width = 6.69 in]{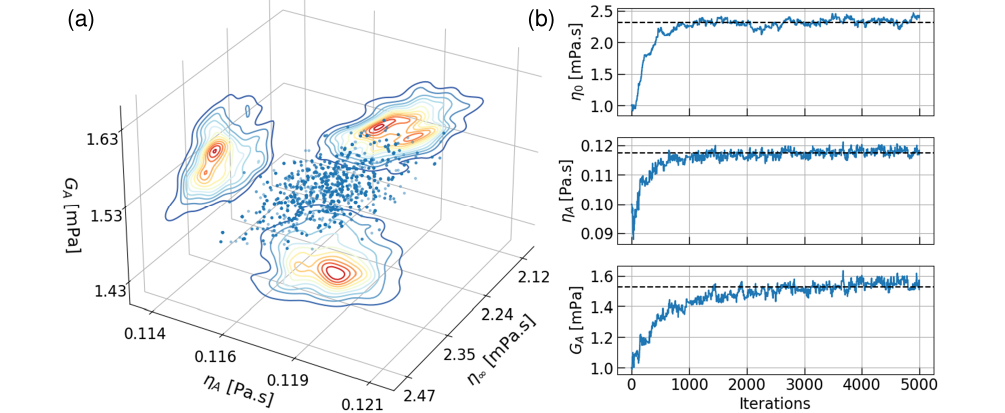}
\vspace{-3mm}
\caption{\small (a) Accepted samples (blue dots) from a Markov chain for the fitting of a kaolinite suspension of $\phi=0.05$, in a three-dimensional subspace of $\eta_\infty$, $\eta_A$, and $G_A$ of the five-dimensional parametric space. Projections (colormap) are marginal density functions. (b) Evolution of the parameters $\eta_\infty$, $\eta_A$, and $G_A$ for a kaolinite suspension of $\phi=0.05$ as a function of iteration number. The Markov chain becomes stationary after around 2000 iterations.}
\label{fig.phase_space}
\end{figure}

To visualize the aforementioned MCMC method in a simple manner, a graphical representation of M-H algorithm is shown in Fig. \ref{fig.MCMC}. Starting with the initial guess $\vec{\theta_0}$ (gray dot), the current sample $\vec{\theta}$ performs a random walk to a new location $\vec{\theta^\prime}$. Samples with low $\alpha$ are rejected (red dots marked with X), while samples with high $\alpha$ are accepted and recorded (blue dots). The Markov chain composed of all accepted samples will slowly drift towards the MLE that maximizes the posterior probability $\mathbb{P}(\vec{\theta}\vert\vec{\mathcal{D}})$, or global minimum in terms of the loss function of the predicted variables (green dot). The Markov chain will then fluctuate around the MLE and reach a stationary distribution. One can use the recorded samples from a stationary Markov chain to approximate the posterior distribution $\mathbb{P}(\vec{\theta}\vert\vec{\mathcal{D}})$. Figure \ref{fig.phase_space}(a) shows an example of a stationary Markov chain in a three-dimensional subspace of $\eta_\infty$, $\eta_A$, and $G_A$, for the model fitting of a kaolinite suspension at $\phi=0.05$. The recorded samples (blue dots) are used to generate marginal probability densities (colormap). Figure \ref{fig.phase_space}(b) shows the evolution of parameters $\eta_\infty$, $\eta_A$, and $G_A$ as a function of the number of iterations for the same kaolinite suspension at $\phi=0.05$, starting with an initial guess of an order of magnitude estimation given as $\vec{\theta_0}=[\eta_\infty,\eta_0,G_A]=[10^{-3}~\mathrm{Pa.s},10^{-1}~\mathrm{Pa.s},10^{-3}~\mathrm{Pa}]$, the Markov chain becomes stationary in approximately 2000 iterations. Samples taken before the Markov chain is stationary are discarded. Only the samples in stationary Markov chain are used for the model prediction shown here.

\section{Results and Discussions}

\subsection{Model parameter fitting results} \label{sec. fitting}

\begin{figure}[t!]
\centering
\includegraphics[width = 3.37 in]{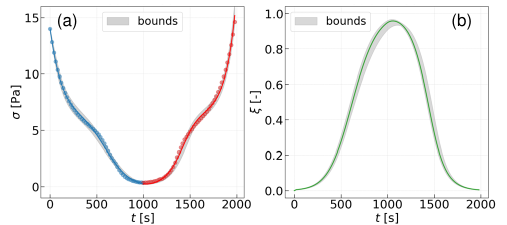}
\vspace{-3mm}
\caption{\small Experimental data (dots) and model fittings (solid curves) of (a) stress ($\sigma$) and (b) structural parameter ($\xi$) as a function of time for a kaolinite suspension at $\phi=0.19$. Gray shades are uncertainty bounds of the fitting results.}
\label{fig. bounds}
\end{figure}

We first discuss the fitting results of the five model parameters that provide estimates of the rheological properties of kaolinite clay suspensions. Table \ref{tab_fit_para} summarizes the values of the five model parameters: $\eta_\infty$, $\eta_A$, $G_A$, $k_A$, $k_R$ (see Sec. \ref{sec. model}), for different kaolinite volume fractions $\phi_k$ and sweep time $\Delta t$. Also listed are the critical shear rate $\dot{\gamma}_c$ and yield stress $\sigma_y$ values which are discussed in Sec. \ref{sec. yield}. Since MCMC method provides the probability distributions of model parameters, we report  both the mean and the standard deviation in Table \ref{tab_fit_para}. Figures \ref{fig. bounds}(a) and \ref{fig. bounds}(b) show the uncertainty bounds for the model fittings of the stress ($\sigma$) and the structural parameter ($\xi$), respectively.

\begin{table}[t!]
\caption{Fitting results of the five model parameters: $\eta_\infty$, $\eta_A$, $G_A$, $k_A$, $k_R$; also listed are critical shear rate $\dot{\gamma}_c$ and yield stress $\sigma_y$. All results are reported with 3 significant figures.}\label{tab_fit_para}
\centering
\begin{tabular}{cccccccc}
\hline\hline
$\phi_k$ & $\eta_\infty\times10^{3}$ & $\eta_A$ & $G_A\times10^{3}$ & $k_A\times10^{3}$ & $k_R\times10^3$ & $\dot{\gamma}_c = k_A/k_R$ & $\sigma_y$\\
 & [Pa.s]& [Pa.s] & [Pa] & [s$^{-1}$] & [--] & [s${}^{-1}$] & [Pa]\\ \hline
$\Delta t=10~$s\\ \hline
0.05 & $2.33\pm0.05$ & $0.114\pm0.002$ & $1.53\pm0.08$ & $17.1\pm1.1$ & $4.03\pm0.28$ & $4.31\pm0.09$ & $0.481\pm0.007$\\
0.12 & $4.26\pm0.32$ & $0.451\pm0.006$ & $11.6\pm0.7$ & $18.7\pm5.6$ & $2.14\pm0.19$ & $8.74\pm0.24$ & $3.90\pm0.08$\\
0.19 & $11.4\pm0.4$ & $1.82\pm0.02$ & $36.8\pm0.9$ & $11.2\pm0.9$ & $3.30\pm0.25$ & $3.36\pm0.05$ & $6.07\pm0.05$\\
0.26 & $131\pm11$ & $2.48\pm0.02$ & $1650\pm260$ & $11.7\pm0.8$ & $0.629\pm0.041$ & $18.6\pm1.8$  & $43.7\pm0.3$\\ \hline
$\Delta t=20~$s\\ \hline
0.05 & $2.27\pm0.06$ & $0.091\pm0.001$ & $0.84\pm0.02$ & $17.0\pm0.8$ & $2.80\pm0.14$ & $6.07\pm0.12$ & $0.538\pm0.008$\\
0.12 & $3.78\pm0.29$ & $0.405\pm0.004$ & $6.72\pm0.13$ & $8.02\pm0.63$ & $0.751\pm0.082$ & $10.7\pm0.3$& $4.31\pm0.08$\\
0.19 & $9.11\pm0.07$ & $1.63\pm0.02$ & $19.6\pm0.8$ & $7.19\pm0.31$ & $1.83\pm0.07$ & $3.93\pm0.04$ & $6.39\pm0.02$\\
0.26 & $122\pm10$ & $2.26\pm0.03$ & $1260\pm360$ & $6.42\pm0.42$ & $0.310\pm0.022$ & $20.7\pm2.5$ & $44.2\pm0.5$\\ \hline\hline
\end{tabular}
\end{table}

For all the $\phi_k$ values, the fitting parameter, the infinite-shear viscosity ($\eta_\infty$) for kaolinite clay suspensions, is much larger than the viscosity of water ($\eta_\mathrm{w}=10^{-3}$~Pa.s). For example, the kaolinite suspension of the lowest volume fraction ($\phi=0.05$) has a infinite-shear viscosity of $\eta_\infty \approx 2\eta_w$. The structural viscosity ($\eta_A$) of the clay suspensions is approximately 50 to 200 times the $\eta_\infty$. Both $\eta_\infty$ and $\eta_A$ increase nonlinearly with $\phi_k$. As the sweep time increases from $\Delta t=10$~s to 20~s, we find that the values of both the viscosities $\eta_\infty$ and $\eta_A$ decrease slightly by $\sim5\%$ for all $\phi_k$ values. This is the result of the suspension exhibiting thixotropy \cite{Mewis_ACIS_2009,Larson_JOR_2015,Larson_JOR_2019}, where the viscosity decreases as the suspension is sheared for a longer time.

Next, we find that the kinetic rate constants $k_R$ and $k_A$ are (on average) reduced by half (except for $\phi=$0.05 for $k_R$) for all kaolinite clay suspensions as $\Delta t$ increases from 10~s to 20~s. Nonetheless, the ratio between these two rate constants, $\dot{\gamma}_c = k_A/k_R$, increases by approximately 15\%. The physical meaning of the ratio $\dot{\gamma}_c$ is discussed in Sec. \ref{sec. yield}. The changes in $k_R$ and $k_A$ with different $\Delta t$ reveal that: (i) kaolinite clay suspensions have multiple time scales corresponding to the relaxation of microstructure; and (ii) the model captures only the dominant time scales corresponding to aging ($k_A^{-1}$) and rejuvenation ($\vert k_R\dot{\gamma}\vert^{-1}$). As $\Delta t$ increase from 10~s to 20~s, the dominant time scales increase and the values of $k_A$ and $k_R$ decrease. Moreover, we find that the values of elastic moduli ($G_A$) are halved, on average, as $\Delta t$ is doubled. This indicates that kaolinite clay suspensions have more than a single stress relaxation time scale, $\lambda=\eta_A/G_A$; and the value of $\lambda$ for $\Delta t=20~$s is almost twice as long as that for $\Delta t=10~$s since $\eta_A$ remains roughly a constant as $\Delta t$ increases (see Table \ref{tab_time_scale}). The multi-timescale nature of kaolinite clay suspensions, as well as the changes in these time scales with $\phi_k$ are discussed in more detail in Sec. \ref{sec. time scale}.

\subsection{Yield stress and rheological hysteresis behaviors}\label{sec. yield}

\begin{figure}[t!]
\centering
\includegraphics[width = 6.69 in]{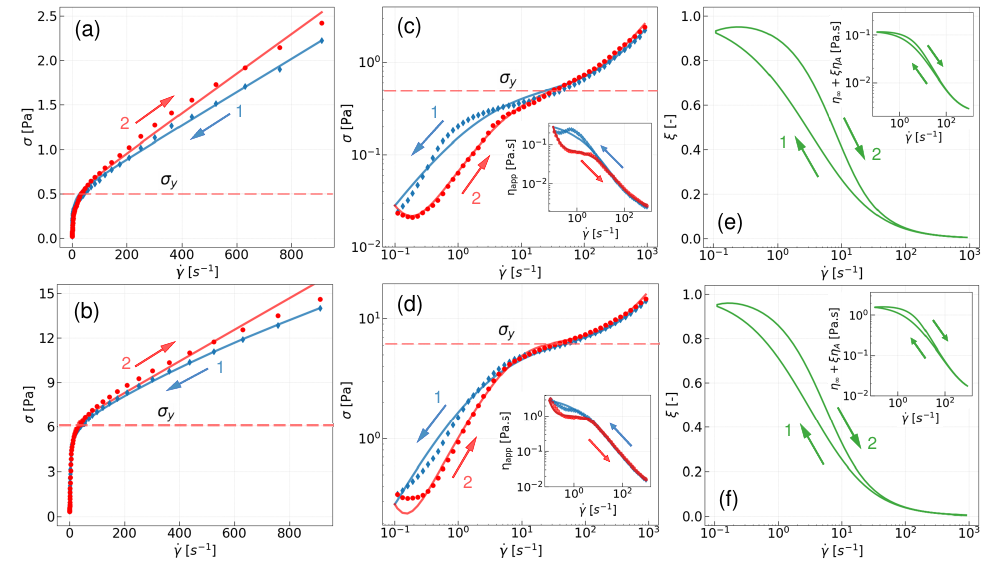}
\vspace{-3mm}
\caption{\small (a-b) Stress $\sigma$ \textit{vs.} shear rate $\dot{\gamma}$ flow curves in linear-linear scale, for kaolinite suspensions of different volume fractions: (a) $\phi=0.05$ and (b) $\phi=0.19$. Arrows indicate the direction of the shear rate sweep: first a downsweep (blue, ``1''), followed by an upsweep (red, ``2''). Symbols are experimental data; solid curves are fittings of the rheological model. Yield stress $\sigma_y$ is indicated by red dashed lines. (c-d) The same flow curves as in (a) and (b), but in log-log scale, for (c) $\phi=0.05$ and (d) $\phi=0.19$. Insets: apparent viscosity, $\eta_\mathrm{app}=\sigma/\dot{\gamma}$, of the same flow curves. (e-f) structural parameter $\xi$ as a function of $\dot{\gamma}$, obtained from the model fittings, for (e) $\phi=0.05$ and (f) $\phi=0.19$. Insets: actual viscosity of the same data, $\eta=\eta_\infty+\xi\eta_A$.}
\label{fig.fittings}
\end{figure}

In this section, we focus on the yield stress and rheological hysteresis behaviors exhibited in the stress ($\sigma$) \textit{vs.} shear rate ($\dot{\gamma}$) flow curves, and how well our model captures these behaviors. Figures \ref{fig.fittings}(a) and \ref{fig.fittings}(b) show the experimental data and model fitting results of $\sigma$-$\dot{\gamma}$ flow curves in linear-linear scale, for kaolinite suspensions of $\phi=0.05$ and 0.19, respectively. We find that the stress increases rapidly with the shear rate until it reaches a critical stress value $\sigma_y$; above $\sigma_y$, the stress exhibits a power-law-like relation with the shear rate. These are typical behaviors of a yield stress fluid, which are captured by the model fitting. Unlike equations of the Herschel-Bulkley form \cite{Herschel_Bulkley_1926}: $\sigma = \sigma_{0} + K\dot{\gamma}^n$, our model does not include an explicit yield stress $\sigma_0$ in the equation of stress [ Equation \eqref{eqn Blackwell}]. Rather, the model captures the yield stress behaviors through a dramatic decrease in viscosity from $(\eta_A+\eta_\infty)$ to $\eta_\infty$ with increasing shear rate. Nevertheless, one can estimate a yield stress from the model as: $\sigma_y=(\eta_A-\eta_\infty)\dot{\gamma}_c$, where $\dot{\gamma}_c=k_A/k_R$ is a critical shear rate at which the material starts to fluidize \cite{Blackwell_JNNFM_2014}. We can see that the model predicted yield stress $\sigma_y$ is located relatively accurately at the yield transition of the flow curves for both $\phi_k$ [red dashed lines in Fig. \ref{fig.fittings}(a) and \ref{fig.fittings}(b)]. Above yield, the stress during upsweep (red) overshoots and stays higher than the stress during downsweep (blue). This rheological hysteresis behavior is also captured by the model. Table \ref{tab_fit_para} lists the values of the yield stress $\sigma_y$ for different volume fraction $\phi_k$ and sweep time $\Delta t$. We find that $\sigma_y$ increases nonlinearly with the volume fraction $\phi_k$, since the additional kaolinite particles increase the viscosities $\eta_A$ and $\eta_\infty$. As $\Delta t$ is doubled, $\sigma_y$ stays roughly the same (Table \ref{tab_fit_para}), which suggests the yield stress of kaolinite clay suspensions is approximately time-independent.

Figures \ref{fig.fittings}(c) and \ref{fig.fittings}(d) highlights the rheological behaviors of kaolinite suspensions below yield stress, by showing the same flow curves as in Fig. \ref{fig.fittings}(a) and \ref{fig.fittings}(b) but in log-log scale. At low shear rate, we find that the stress during upsweep (red) undershoots and stays below that during downsweep (blue), until it reaches the yield stress. This creates a hysteresis loop below yield that is inverted compared to the flow curves above yield, in which case the stress overshoots during upsweep. Our model captures this double-loop hysteresis behaviors quite well. Similar double-loop hysteresis behaviors have been observed for other structured fluids in previous studies \cite{Potato_CP_2016,Fielding_SM_2017}. And it has been proposed that the looping behavior below yield is the signature of viscoelasticity \cite{Fielding_SM_2017}: the elastic stress initially accumulated at the high shear rates continues to relax during upsweep, which leads to lower stress values during upsweep than downsweep. This mechanism requires the time scale of stress relaxation to be comparable to the time scale of microstructural relaxation; and we shall see that this is the case for kaolinite suspensions at these volume fractions in Sec. \ref{sec. time scale}. Overall, these results suggest that the behaviors of kaolinite clay suspensions are primarily viscoelastic below yield transition, which is consistent with previous modelling of structured fluids \cite{DeSouza_JNNFM_2009,DeSouza_SoftMatt_2011}.

Insets of Figs. \ref{fig.fittings}(c) and \ref{fig.fittings}(d) show the apparent viscosity, $\eta_\mathrm{app}=\sigma/\dot{\gamma}$, of the same flow curves. We find that $\eta_\mathrm{app}$ increases non-monotonically during downsweep (blue), which is likely due to the coexistence of multiple stress relaxation times. This non-monotonic behavior is not captured by our model, since our model only has a single stress relaxation time $\lambda$. During upsweep (red), the model fits the flow curves relatively better than downsweep. This can be understood as follows. Since elastic stress is mainly accumulated at the start of downsweep, faster stress relaxation modes have already fully relaxed at the end of downsweep. And our model captures the longest relaxation time $\lambda$ associated with the slowest stress relaxation mode during upsweep. Another limitation of the model is that it only has a single microstructural relaxation time $k_A^{-1}$, thus it fails to predict non-monotonic microstructural relaxation behaviors \cite{Joshi_JOR_2022}. Both limitations of our model can be responsible for the discrepancy between the experimental data and the model fittings. These results suggest that structural kinetic models with multiple relaxation time scales may better capture the rheological behavior of clay suspensions, especially the non-monotonic relaxation behaviors in Fig. \ref{fig.fittings}(c) and \ref{fig.fittings}(d). It has been shown that viscoelastic models with multiple stress relaxation times can capture non-monotonic stress relaxation behaviors, examples including the soft glassy rheology (SGR) model \cite{SGR_PRL_1997,SGR_PRE_1998,Fielding_SM_2017} and the mode coupling theory (MCT) model \cite{Larson_rheology_1999,Joshi_SM_2015}. It has also been shown that model with multiple microstructural relaxation time scales are able to capture non-monotonic thixotropic relaxation behaviors, such as the multimode structural kinetics model \cite{Larson_JOR_2016,Larson_JOR_2018}.

Figures \ref{fig.fittings}(e) and \ref{fig.fittings}(f) show the structural parameter $\xi$ as a function of shear rate $\dot{\gamma}$, for kaolinite suspensions of $\phi_k=0.05$ and 0.19. Since $\xi$ is unobservable, there is no experimental data shown in the figures. We find that for all shear rates, the structural parameter during upsweep (``2'') is always larger than that during downsweep (``1''), leading to hysteresis loops in the $\xi$-$\dot{\gamma}$ flow curves. Since the fluid viscosity is directly related to the structural parameter by $\eta=\eta_\infty+\xi\eta_A$, the viscosity during the upsweep is also always higher than that during downsweep, as shown in the insets of Fig. \ref{fig.fittings}(e) and \ref{fig.fittings}(f). This is expected since downsweep has a much stronger shear history than upsweep: the high $\dot{\gamma}$ values at the start of downsweep destroy most microstructure in the fluid, while the low $\dot{\gamma}$ values at the beginning of upsweep allow more microstructure to build up. These results provide further evidence that the hysteresis behaviors in Figs. \ref{fig.fittings}(c) and \ref{fig.fittings}(d) are caused by viscoelasticity. The fluid has a lower stress value despite having a higher viscosity during upsweep (than downsweep), which indicates the looping behaviors below yield stress are the results of viscoelasticity, instead of the change in viscosity due to thixotropy.

\subsection{Stress \textit{vs.} microstructural relaxation time scales} \label{sec. time scale}

Kaolinite suspensions exhibit thixotropy since kaolinite clay particles form microstructure in the suspensions that physically breaks down and builds up over time \cite{Rand_JCIS_1977,Gupta_JCIS_Struc_2011,Phan_Thien_Langmuir_2015,Neelakantan_Mineral_2018}. In our model, there are two time scales associated with thixotropy: $k_A^{-1}$ for the buildup (or relaxation) of microstructure, and $\vert k_R\dot{\gamma}\vert^{-1}$ for the breakdown of microstructure. Despite the former is often termed ``thixotropic time scale'' in the literature \cite{Beris_JNNFM_2002,Larson_JOR_2019,Jamali_JoR_2022}, here we refer to $k_A^{-1}$ as the ``microstructural relaxation time scale'' to distinguish between the two different time scales associated with thixotropy. In this section, we focus on the comparison of two relaxation time scales: $\lambda$ associated with the relaxation of elastic stress, and $k_A^{-1}$ associated with the relaxation of microstructure in the suspension. The ratio between the two time scales, $\lambda k_A$, provides a possible classification of thixotropy into ``ideal'' and ``viscoelastic'' \cite{Mewis_ACIS_2009,Larson_JOR_2015,Larson_JOR_2019}. ``Ideal'' thixotropic materials have a much shorter relaxation time of elastic stress than that of microstructure, and $\lambda k_A\to0$; ``viscoelastic'' thixotropic materials have a stress relaxation time that is comparable to the microstructural relaxation time, and $\lambda k_A\sim\mathcal{O}(10^0)$.

\begin{table}[t!]
\caption{Stress relaxation time scale $\lambda$, microstructural relaxation time scale $k_A^{-1}$, and the ratio between the 2 time scales $\lambda k_A$.}\label{tab_time_scale}
\centering
\begin{tabular}{cccc}
\hline\hline
$\phi_k$ & $\lambda$ [s] & $k_A^{-1}$ [s] & $\lambda k_A$ [--]\\ \hline
$\Delta t=10~$s\\ \hline
0.05 & $77.0\pm5.4$ & $58.7\pm3.7$ & $1.32\pm0.07$\\
0.12 & $38.8\pm1.4$ & $54.0\pm5.6$ & $0.73\pm0.06$\\
0.19 & $49.5\pm3.4$ & $93.1\pm4.5$ & $0.56\pm0.02$\\
0.26 & $1.5\pm0.4$ & $85.4\pm6.4$ & $0.02\pm0.00$\\ \hline
$\Delta t=20~$s\\ \hline
0.05 & $107.3\pm1.7$ & $59.1\pm2.8$ & $1.82\pm0.08$\\
0.12 & $60.8\pm2.7$ & $125.8\pm6.5$ & $0.49\pm0.05$\\
0.19 & $83.1\pm3.7$ & $139.4\pm5.8$ & $0.60\pm0.01$\\
0.26 & $1.8\pm0.6$ & $156.2\pm7.2$ & $0.01\pm0.00$\\ \hline\hline
\end{tabular}
\end{table}

\begin{figure}[t!]
\centering
\includegraphics[width=3.37in]{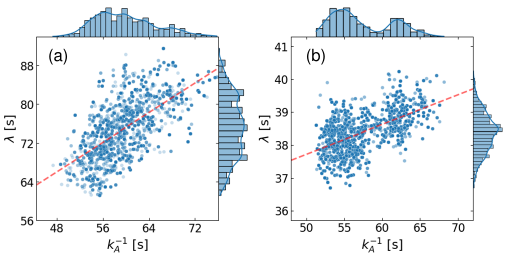}
\vspace{-3mm}
\caption{\small (a) Examples of kaolinite suspensions possessing linearly correlated spectra of stress relaxation time scale $\lambda$ and microstructural relaxation time scale $k_A^{-1}$, for $\phi_k=0.05$ and $\Delta t=10~$s. (b) Examples of kaolinite suspensions possessing a single dominant stress relaxation time scale $\lambda$, and multiple dominant microstructural relaxation time scales $k_A^{-1}$, for $\phi_k=0.12$ and $\Delta t=10~$s. Here, joint distributions are illustrated by scatter plots of the accepted samples from the Markov chain (blue dots), and marginal distributions are shown by the histograms on the side (bar plots); red dashed lines are linear regressions of the data.}
\label{fig.joint}
\end{figure}

Table \ref{tab_time_scale} lists the fitting results of the stress relaxation time $\lambda$, the microstructural relaxation time $k_A^{-1}$, and the ratio between the two time scales $\lambda k_A$, for different volume fractions $\phi_k$ and sweep time $\Delta t$. We first discuss the changes in these time scales with $\phi_k$. We find that both time scales are at the order of 30 seconds to several minutes (except for $\phi_k = 0.26$, which is discussed later). The stress relaxation time $\lambda$ decreases with increasing $\phi_k$, while the microstructural relaxation time $k_A^{-1}$ increases with $\phi_k$. As a result, the time scale ratio $\lambda k_A$ decreases with $\phi_k$, indicating that the kaolinite suspensions become more thixotropic and less viscoelastic. An exception is for $\Delta t=20~$s, at $\phi_k=0.12$ and $\phi_k=0.19$. We believe that this is due to the multi-timescale nature of kaolinite clay suspension, which will be explained later in the discussion of Fig. \ref{fig.joint}. At relatively low volume fractions ($\phi_k < 0.26$), $\lambda k_A\sim\mathcal{O}(10^0)$. At the highest volume fraction ($\phi_k = 0.26$), however, $\lambda k_A$ decreases dramatically to $\sim\mathcal{O}(10^{-2})$, which indicates the kaolinite suspension is almost ideally thixotropic with negligible viscoelasticity. The changes in the rheological behaviors of kaolinite suspensions can be explained as follows. Kaolinite particles are charged platelets; the faces and edges of kaolinite particles carry opposite charges. As $\phi_k$ increases, the dominant interaction between kaolinite particles transitions from an edge-face attraction to a face-face repulsion, due to steric effects \cite{Phan_Thien_Langmuir_2015}. This transition from attractive to repulsive interaction will reduce the viscoelastic behaviors of kaolinite suspensions \cite{chattopadhyay2022effect}, which is responsible for the dramatic decreases in $\lambda$ (from $\sim30$~s to $\sim1$~s) and $\lambda k_A$ (from $\sim0.5$ to $\sim0.01$) at the highest $\phi_k$.

Next, we focus on the changes in stress and microstructural relaxation times with different sweep time $\Delta t$. We find that both time scales $\lambda$ and $k_A^{-1}$ increase as $\Delta t$ is doubled from 10~s to 20~s (by an average of 53\% and 68\%, respectively). As discussed in Sec. \ref{sec. fitting}, this is a result of the multi-timescale nature of kaolinite clay suspensions, as they possess multiple stress and microstructural relaxation times. Since the model used here only has a single stress relaxation time $\lambda$ and a single microstructural relaxation time $k_A^{-1}$, it captures only the dominant time scales of the relaxation processes. As $\Delta t$ increases, the dominant time scales captured by the model will increase accordingly, which is shown by the increase in stress and microstructural relaxation times with $\Delta t$ (Table \ref{tab_time_scale}). Since $\lambda$ and $k_A^{-1}$ increase by roughly the same percentage, the ratio between them $\lambda k_A^{-1}$ remains similar as $\Delta t$ is increased.

Another piece of evidence of the multi-timescale nature of kaolinite suspensions can be shown by the probability distribution phase of the relaxation time scales. Figure \ref{fig.joint}(a) shows the joint probability distribution of $\lambda$ and $k_A^{-1}$ for $\phi=0.05$ and $\Delta t=10~$s. The marginal distributions (bar plots) of $\lambda$ and $k_A^{-1}$ show that both time scales vary by approximately $\pm 30$~s, and there is no evident modal time scale (peak) in the distributions, indicating the existence of a spectrum of time scales. Interestingly, the joint distribution exhibits a linear correlation between $\lambda$ and $k_A^{-1}$, which suggests we can have a relatively accurate estimate of the intrinsic time scale ratio $\lambda k_A$, despite not having accurate measures of $\lambda$ and $k_A^{-1}$ separately. Figure \ref{fig.joint}(b) shows the joint probability distribution of $\lambda$ and $k_A^{-1}$ for $\phi=0.12$ and $\Delta t=10~$s. The marginal distribution of $k_A^{-1}$ shows two modal time scales, peaked at 54~s and 62~s, respectively. Such bimodal distribution of $k_A^{-1}$ can sometime lead to under- or over-estimation of relaxation time scale, which is likely responsible for the unexpected increase in $\lambda k_A$ from $\phi_k=0.12$ to $\phi_k=0.19$ for $\Delta t=20~$s. The marginal distribution of $\lambda$ shows a single modal time scale at approximately $38.5~$s, with narrow tails of a standard deviation of $\sim1~$s. The joint distribution of $\lambda$ and $k_A^{-1}$ still demonstrates a weak linear correlation, but not as evident as in Fig. \ref{fig.joint}(a). Both figures show that kaolinite clay suspensions possess multiple microstructural and stress relaxation times. Conventionally, the material's intrinsic time scales are obtained by plotting the hysteresis area ($A_\sigma$) as a function of the sweep time ($\Delta t$) and identifying the peak in the $A_\sigma$--$\Delta t$ plot \cite{Divoux_PRL_2013,Mckinley_PRL_2019,Jamali_JoR_2022}. This procedure requires performing experiments with a range of sweep times; and the effects of thixotropy and viscoelasticity on the identified time scales can be challenging to separate \cite{Joshi_ArXiv_2022}. Here, the reported Bayesian method provides estimates of the material’s intrinsic time scales by directly fitting to the flow curves, without the need for performing experiments of different sweep times.

\subsection{Gravitational aging and microstructural relaxation} \label{sec. gravity}
We now discuss the origin of ``aging'', i.e., the recovery of microstructure in kaolinite clay suspensions. In many colloidal suspensions, the recovery of microstructure is considered to be induced by Brownian motion \cite{Mewis_JOR_2005,Mewis_ACIS_2009,Joshi_RA_2018,Larson_JOR_2018,Larson_JOR_2019,Larson_SoftMatt_2019}. We shall see that this may not be the case for kaolinite clay suspensions, since kaolinite particles are essentially non-Brownian. The extent of non-thermal nature of the constituent particles is estimated using the gravitational P\'eclet number, which is the ratio between the time scale of Brownian motion and the sedimentation time scale \cite{Weeks_RPP_2012,Pradeep_SM_2020}:
\begin{equation}
    Pe_g=\frac{4\pi r^4\Delta\rho g}{3k_BT},
\end{equation}
where $r$ is the average particle radius, $\Delta\rho$ is the density difference between kaolinite particles and water, $g$ is the gravitational acceleration, $k_B$ is the Boltzmann constant, and $T$ is the Kelvin temperature. Kaolinite particles have a modal nominal diameter of $d\approx10~\umu$m \cite{Seiphoori_GRL_2021,Phan_Thien_Langmuir_2015}, or a particle radius of $r\approx5~\umu$m. This leads to a gravitational P\'eclet number of $Pe_g\sim\mathcal{O}(10^4)$, which indicates that kaolinite particles are non-Brownian in nature and aging in kaolinite suspensions is mainly driven by gravitational settling, rather than Brownian motion. The aging process in kaolinite suspensions can be understood as follows. As kaolinite particles sediment, they pile up onto each other, and aggregate due to electrostatic forces, which constantly generates new microstructures in the suspensions. We should notice that the system does not restore its original state by forming new microstructure. Therefore, we prefer the terminology ``restructuring'', instead of ``recovery'' of microstructure.

\begin{figure}[t!]
\centering
\includegraphics[width=3.37in]{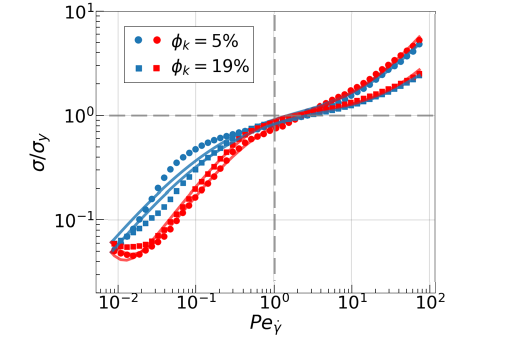}
\vspace{-3mm}
\caption{\small Normalized stress $\sigma/\sigma_y$ \textit{vs.} the shear P\'eclet number $Pe_{\dot{\gamma}}=9\eta_\mathrm{w}\dot{\gamma}/(2\Delta\rho gr)$. We find that the yield stress $\sigma_y$ is attained approximately at $Pe_{\dot{\gamma}}\sim1$.}
\label{fig.Pe}
\end{figure}

This hypothesis allows us to non-dimensionalize our $\sigma$-$\dot{\gamma}$ flow curves. To re-scale our shear rate, we estimate the shear P\'eclet number $Pe_{\dot{\gamma}}$, which is the ratio of the sedimentation time scale and the time scale of the fluid shear (or experimental time scale):
\begin{equation}
    Pe_{\dot{\gamma}}=\frac{9\eta_\mathrm{w}\dot{\gamma}}{2\Delta\rho gr}.
\end{equation}
Figure \ref{fig.Pe} shows the normalized stress, scaled using inherent yield stress scale, $\sigma/\sigma_y$ as a function of the shear P\'eclet number $Pe_{\dot{\gamma}}$. We find that in the re-scaled flow curves, the yield stress $\sigma_y$ matches approximately at $Pe_{\dot{\gamma}}=1$. Below yield stress, $Pe_{\dot{\gamma}}<1$, the fluid shear is too weak to overrun the gravitational aging. Above yield stress, $Pe_{\dot{\gamma}}>1$, the shear flow is strong enough to rejuvenate the sample from aging, and the suspension starts to fluidize. The yield stress of kaolinite suspensions is a result of the competition between two time scales: the time scale of fluid shear associated with the breakdown of microstructure, and the time scale of gravitational settling associated with the buildup of microstructure. We note that the simple scaling of $Pe_{\dot{\gamma}}$ considers neither the electrostatic force nor the interaction between kaolinite clay particles. Nonetheless, this limitation does not prevent the scaling from qualitatively capturing the yield transition of kaolinite clay suspensions.

\section{Conclusions}
To summarize, we experimentally examine the rheology of kaolinite clay suspensions using a shear rate sweep protocol consisted of consecutive downsweep and upsweep; the flow curves show both yield stress and rheological hysteresis behaviors for various kaolinite volume fractions. These behaviors are captured by a microstructural viscoelastic model using Markov chain Monte Carlo (MCMC) fitting method. The method also allow us to estimate the rheological properties of the clay suspensions, such as viscosities, stress and microstructural relaxation time scales, and the yield stress. The comparison of stress relaxation time $\lambda$ and microstructural relaxation time $k_A^{-1}$ suggests that kaolinite clay suspensions are strongly viscoelastic and weakly thixotropic at relatively low volume fractions ($\phi_k<0.26$), while being almost inelastic and purely thixotropic at the highest volume fractions ($\phi_k=0.26$). We have also shown that kaolinite suspensions possess multiple or a spectrum of stress and microstructural relaxation time scales. Overall, these results can provide insights for developing predictive models of natural materials like mud and clay, as well as other yield stress materials and known structured fluids.

\begin{acknowledgments}
We thank Yogesh Joshi, Philippe Coussot, Paris Perdikaris, Georgios Kissas, Yibo Yang, and Larry Galloway for helpful discussions. This work was supported by the National Science Foundation (NSF) Grant DMR-1709763 (R.R., P.E.A.), and by the Army Research Office (ARO) Grant W911-NF-16-1-0290 (S.K.A., D.J.J. and P.E.A.), and by ARO Grant W911-NF-20-1-0113 (S.P., D.J.J.).
\end{acknowledgments}

\bibliography{reference}

\begin{thebibliography}{95}%
\makeatletter
\providecommand \@ifxundefined [1]{%
 \@ifx{#1\undefined}
}%
\providecommand \@ifnum [1]{%
 \ifnum #1\expandafter \@firstoftwo
 \else \expandafter \@secondoftwo
 \fi
}%
\providecommand \@ifx [1]{%
 \ifx #1\expandafter \@firstoftwo
 \else \expandafter \@secondoftwo
 \fi
}%
\providecommand \natexlab [1]{#1}%
\providecommand \enquote  [1]{``#1''}%
\providecommand \bibnamefont  [1]{#1}%
\providecommand \bibfnamefont [1]{#1}%
\providecommand \citenamefont [1]{#1}%
\providecommand \href@noop [0]{\@secondoftwo}%
\providecommand \href [0]{\begingroup \@sanitize@url \@href}%
\providecommand \@href[1]{\@@startlink{#1}\@@href}%
\providecommand \@@href[1]{\endgroup#1\@@endlink}%
\providecommand \@sanitize@url [0]{\catcode `\\12\catcode `\$12\catcode
  `\&12\catcode `\#12\catcode `\^12\catcode `\_12\catcode `\%12\relax}%
\providecommand \@@startlink[1]{}%
\providecommand \@@endlink[0]{}%
\providecommand \url  [0]{\begingroup\@sanitize@url \@url }%
\providecommand \@url [1]{\endgroup\@href {#1}{\urlprefix }}%
\providecommand \urlprefix  [0]{URL }%
\providecommand \Eprint [0]{\href }%
\providecommand \doibase [0]{https://doi.org/}%
\providecommand \selectlanguage [0]{\@gobble}%
\providecommand \bibinfo  [0]{\@secondoftwo}%
\providecommand \bibfield  [0]{\@secondoftwo}%
\providecommand \translation [1]{[#1]}%
\providecommand \BibitemOpen [0]{}%
\providecommand \bibitemStop [0]{}%
\providecommand \bibitemNoStop [0]{.\EOS\space}%
\providecommand \EOS [0]{\spacefactor3000\relax}%
\providecommand \BibitemShut  [1]{\csname bibitem#1\endcsname}%
\let\auto@bib@innerbib\@empty
\bibitem [{\citenamefont {Hungr}\ \emph {et~al.}(2014)\citenamefont {Hungr},
  \citenamefont {Leroueil},\ and\ \citenamefont
  {Picarelli}}]{Hungr_Landslides_2014}%
  \BibitemOpen
  \bibfield  {author} {\bibinfo {author} {\bibfnamefont {O.}~\bibnamefont
  {Hungr}}, \bibinfo {author} {\bibfnamefont {S.}~\bibnamefont {Leroueil}},\
  and\ \bibinfo {author} {\bibfnamefont {L.}~\bibnamefont {Picarelli}},\
  }\bibfield  {title} {\bibinfo {title} {The varnes classification of landslide
  types, an update},\ }\href {https://doi.org/10.1007/s10346-013-0436-y}
  {\bibfield  {journal} {\bibinfo  {journal} {Landslides}\ }\textbf {\bibinfo
  {volume} {11}},\ \bibinfo {pages} {167} (\bibinfo {year} {2014})}\BibitemShut
  {NoStop}%
\bibitem [{\citenamefont {Pimentel}(2006)}]{Pimentel_Soil_2006}%
  \BibitemOpen
  \bibfield  {author} {\bibinfo {author} {\bibfnamefont {D.}~\bibnamefont
  {Pimentel}},\ }\bibfield  {title} {\bibinfo {title} {Soil erosion: A food and
  environmental threat},\ }\href {https://doi.org/10.1007/s10668-005-1262-8}
  {\bibfield  {journal} {\bibinfo  {journal} {Environ. Dev. Sustain.}\ }\textbf
  {\bibinfo {volume} {8}},\ \bibinfo {pages} {119} (\bibinfo {year}
  {2006})}\BibitemShut {NoStop}%
\bibitem [{\citenamefont {Dunne}\ and\ \citenamefont
  {Jerolmack}(2020)}]{dunne2020sets}%
  \BibitemOpen
  \bibfield  {author} {\bibinfo {author} {\bibfnamefont {K.~B.}\ \bibnamefont
  {Dunne}}\ and\ \bibinfo {author} {\bibfnamefont {D.~J.}\ \bibnamefont
  {Jerolmack}},\ }\bibfield  {title} {\bibinfo {title} {What sets river
  width?},\ }\href {https://doi.org/10.1126/sciadv.abc1505} {\bibfield
  {journal} {\bibinfo  {journal} {Sci. Adv.}\ }\textbf {\bibinfo {volume} {6}}
  (\bibinfo {year} {2020})}\BibitemShut {NoStop}%
\bibitem [{\citenamefont {Phillips}\ \emph {et~al.}(2022)\citenamefont
  {Phillips}, \citenamefont {Masteller}, \citenamefont {Slater}, \citenamefont
  {Dunne}, \citenamefont {Francalanci}, \citenamefont {Lanzoni}, \citenamefont
  {Merritts}, \citenamefont {Lajeunesse},\ and\ \citenamefont
  {Jerolmack}}]{phillips2022threshold}%
  \BibitemOpen
  \bibfield  {author} {\bibinfo {author} {\bibfnamefont {C.~B.}\ \bibnamefont
  {Phillips}}, \bibinfo {author} {\bibfnamefont {C.~C.}\ \bibnamefont
  {Masteller}}, \bibinfo {author} {\bibfnamefont {L.~J.}\ \bibnamefont
  {Slater}}, \bibinfo {author} {\bibfnamefont {K.~B.}\ \bibnamefont {Dunne}},
  \bibinfo {author} {\bibfnamefont {S.}~\bibnamefont {Francalanci}}, \bibinfo
  {author} {\bibfnamefont {S.}~\bibnamefont {Lanzoni}}, \bibinfo {author}
  {\bibfnamefont {D.~J.}\ \bibnamefont {Merritts}}, \bibinfo {author}
  {\bibfnamefont {E.}~\bibnamefont {Lajeunesse}},\ and\ \bibinfo {author}
  {\bibfnamefont {D.~J.}\ \bibnamefont {Jerolmack}},\ }\bibfield  {title}
  {\bibinfo {title} {Threshold constraints on the size, shape and stability of
  alluvial rivers},\ }\href
  {https://www.nature.com/articles/s43017-022-00282-z} {\bibfield  {journal}
  {\bibinfo  {journal} {Nat. Rev. Earth Environ.}\ }\textbf {\bibinfo {volume}
  {3}},\ \bibinfo {pages} {406} (\bibinfo {year} {2022})}\BibitemShut {NoStop}%
\bibitem [{\citenamefont {Barnes}(2010)}]{Barnes_Civil_2000}%
  \BibitemOpen
  \bibfield  {author} {\bibinfo {author} {\bibfnamefont {G.~E.}\ \bibnamefont
  {Barnes}},\ }\href {https://books.google.com/books?id=asdJSAAACAAJ} {\emph
  {\bibinfo {title} {Soil Mechanics: Principles and Practice}}}\ (\bibinfo
  {publisher} {Palgrave Macmillan},\ \bibinfo {year} {2010})\BibitemShut
  {NoStop}%
\bibitem [{\citenamefont {Van~Dyke}(1998)}]{VanDyke_drilling_1998}%
  \BibitemOpen
  \bibfield  {author} {\bibinfo {author} {\bibfnamefont {K.}~\bibnamefont
  {Van~Dyke}},\ }\href {https://books.google.com/books?id=j19QAQAAIAAJ} {\emph
  {\bibinfo {title} {Drilling Fluids, Mud Pumps, and Conditioning Equipment}}}\
  (\bibinfo  {publisher} {University of Texas at Austin, Petroleum Extension
  Service},\ \bibinfo {year} {1998})\BibitemShut {NoStop}%
\bibitem [{\citenamefont {Jerolmack}\ and\ \citenamefont
  {Daniels}(2019)}]{Jerolmack_NRP_2019}%
  \BibitemOpen
  \bibfield  {author} {\bibinfo {author} {\bibfnamefont {D.~J.}\ \bibnamefont
  {Jerolmack}}\ and\ \bibinfo {author} {\bibfnamefont {K.~E.}\ \bibnamefont
  {Daniels}},\ }\bibfield  {title} {\bibinfo {title} {Viewing earth's surface
  as a soft-matter landscape},\ }\href
  {https://doi.org/10.1038/s42254-019-0111-x} {\bibfield  {journal} {\bibinfo
  {journal} {Nat. Rev. Phys.}\ }\textbf {\bibinfo {volume} {1}},\ \bibinfo
  {pages} {716} (\bibinfo {year} {2019})}\BibitemShut {NoStop}%
\bibitem [{\citenamefont {Coussot}\ \emph {et~al.}(2002)\citenamefont
  {Coussot}, \citenamefont {Nguyen}, \citenamefont {Huynh},\ and\ \citenamefont
  {Bonn}}]{Coussot_JOR_2002}%
  \BibitemOpen
  \bibfield  {author} {\bibinfo {author} {\bibfnamefont {P.}~\bibnamefont
  {Coussot}}, \bibinfo {author} {\bibfnamefont {Q.~D.}\ \bibnamefont {Nguyen}},
  \bibinfo {author} {\bibfnamefont {H.~T.}\ \bibnamefont {Huynh}},\ and\
  \bibinfo {author} {\bibfnamefont {D.}~\bibnamefont {Bonn}},\ }\bibfield
  {title} {\bibinfo {title} {Viscosity bifurcation in thixotropic, yielding
  fluids},\ }\href {https://doi.org/10.1122/1.1459447} {\bibfield  {journal}
  {\bibinfo  {journal} {J. Rheol.}\ }\textbf {\bibinfo {volume} {46}},\
  \bibinfo {pages} {573} (\bibinfo {year} {2002})}\BibitemShut {NoStop}%
\bibitem [{\citenamefont {Da~Cruz}\ \emph {et~al.}(2002)\citenamefont
  {Da~Cruz}, \citenamefont {Chevoir}, \citenamefont {Bonn},\ and\ \citenamefont
  {Coussot}}]{Cruz_PRE_2002}%
  \BibitemOpen
  \bibfield  {author} {\bibinfo {author} {\bibfnamefont {F.}~\bibnamefont
  {Da~Cruz}}, \bibinfo {author} {\bibfnamefont {F.}~\bibnamefont {Chevoir}},
  \bibinfo {author} {\bibfnamefont {D.}~\bibnamefont {Bonn}},\ and\ \bibinfo
  {author} {\bibfnamefont {P.}~\bibnamefont {Coussot}},\ }\bibfield  {title}
  {\bibinfo {title} {Viscosity bifurcation in granular materials, foams, and
  emulsions},\ }\href {https://doi.org/10.1103/PhysRevE.66.051305} {\bibfield
  {journal} {\bibinfo  {journal} {Phys. Rev. E}\ }\textbf {\bibinfo {volume}
  {66}},\ \bibinfo {pages} {051305} (\bibinfo {year} {2002})}\BibitemShut
  {NoStop}%
\bibitem [{\citenamefont {Bonn}\ \emph {et~al.}(2017)\citenamefont {Bonn},
  \citenamefont {Denn}, \citenamefont {Berthier}, \citenamefont {Divoux},\ and\
  \citenamefont {Manneville}}]{Bonn_RMP_2017}%
  \BibitemOpen
  \bibfield  {author} {\bibinfo {author} {\bibfnamefont {D.}~\bibnamefont
  {Bonn}}, \bibinfo {author} {\bibfnamefont {M.~M.}\ \bibnamefont {Denn}},
  \bibinfo {author} {\bibfnamefont {L.}~\bibnamefont {Berthier}}, \bibinfo
  {author} {\bibfnamefont {T.}~\bibnamefont {Divoux}},\ and\ \bibinfo {author}
  {\bibfnamefont {S.}~\bibnamefont {Manneville}},\ }\bibfield  {title}
  {\bibinfo {title} {Yield stress materials in soft condensed matter},\ }\href
  {https://doi.org/10.1103/RevModPhys.89.035005} {\bibfield  {journal}
  {\bibinfo  {journal} {Rev. Mod. Phys.}\ }\textbf {\bibinfo {volume} {89}},\
  \bibinfo {pages} {035005} (\bibinfo {year} {2017})}\BibitemShut {NoStop}%
\bibitem [{\citenamefont {Frigaard}(2019)}]{Frigaard_COCIS_2019}%
  \BibitemOpen
  \bibfield  {author} {\bibinfo {author} {\bibfnamefont {I.}~\bibnamefont
  {Frigaard}},\ }\bibfield  {title} {\bibinfo {title} {Simple yield stress
  fluids},\ }\href {https://doi.org/10.1016/j.cocis.2019.03.002} {\bibfield
  {journal} {\bibinfo  {journal} {Curr. Opin. Colloid Interface Sci.}\ }\textbf
  {\bibinfo {volume} {43}},\ \bibinfo {pages} {80} (\bibinfo {year}
  {2019})}\BibitemShut {NoStop}%
\bibitem [{\citenamefont {Coussot}(1995)}]{Coussot_PRL_1995}%
  \BibitemOpen
  \bibfield  {author} {\bibinfo {author} {\bibfnamefont {P.}~\bibnamefont
  {Coussot}},\ }\bibfield  {title} {\bibinfo {title} {Structural similarity and
  transition from newtonian to non-newtonian behavior for clay-water
  suspensions},\ }\href {https://doi.org/10.1103/PhysRevLett.74.3971}
  {\bibfield  {journal} {\bibinfo  {journal} {Phys. Rev. Lett.}\ }\textbf
  {\bibinfo {volume} {74}},\ \bibinfo {pages} {3971} (\bibinfo {year}
  {1995})}\BibitemShut {NoStop}%
\bibitem [{\citenamefont {Jogun}\ and\ \citenamefont
  {Zukoski}(1999)}]{Zukoski_JOR_1999}%
  \BibitemOpen
  \bibfield  {author} {\bibinfo {author} {\bibfnamefont {S.~M.}\ \bibnamefont
  {Jogun}}\ and\ \bibinfo {author} {\bibfnamefont {C.~F.}\ \bibnamefont
  {Zukoski}},\ }\bibfield  {title} {\bibinfo {title} {Rheology and
  microstructure of dense suspensions of plate-shaped colloidal particles},\
  }\href {https://doi.org/10.1122/1.551013} {\bibfield  {journal} {\bibinfo
  {journal} {J. Rheol.}\ }\textbf {\bibinfo {volume} {43}},\ \bibinfo {pages}
  {847} (\bibinfo {year} {1999})}\BibitemShut {NoStop}%
\bibitem [{\citenamefont {Brown}\ \emph {et~al.}(2000)\citenamefont {Brown},
  \citenamefont {Clarke}, \citenamefont {Convert},\ and\ \citenamefont
  {Rennie}}]{Brown_JOR_2000}%
  \BibitemOpen
  \bibfield  {author} {\bibinfo {author} {\bibfnamefont {A.~B.~D.}\
  \bibnamefont {Brown}}, \bibinfo {author} {\bibfnamefont {S.~M.}\ \bibnamefont
  {Clarke}}, \bibinfo {author} {\bibfnamefont {P.}~\bibnamefont {Convert}},\
  and\ \bibinfo {author} {\bibfnamefont {A.~R.}\ \bibnamefont {Rennie}},\
  }\bibfield  {title} {\bibinfo {title} {Orientational order in concentrated
  dispersions of plate-like kaolinite particles under shear},\ }\href
  {https://doi.org/10.1122/1.551093} {\bibfield  {journal} {\bibinfo  {journal}
  {J. Rheol.}\ }\textbf {\bibinfo {volume} {44}},\ \bibinfo {pages} {221}
  (\bibinfo {year} {2000})}\BibitemShut {NoStop}%
\bibitem [{\citenamefont {Zbik}\ \emph {et~al.}(2008)\citenamefont {Zbik},
  \citenamefont {Smart},\ and\ \citenamefont {Morris}}]{Zbik_Kaolin_2008}%
  \BibitemOpen
  \bibfield  {author} {\bibinfo {author} {\bibfnamefont {M.~S.}\ \bibnamefont
  {Zbik}}, \bibinfo {author} {\bibfnamefont {R.~S.}\ \bibnamefont {Smart}},\
  and\ \bibinfo {author} {\bibfnamefont {G.~E.}\ \bibnamefont {Morris}},\
  }\bibfield  {title} {\bibinfo {title} {Kaolinite flocculation structure},\
  }\href {https://doi.org/https://doi.org/10.1016/j.jcis.2008.08.063}
  {\bibfield  {journal} {\bibinfo  {journal} {J. Colloid Interface Sci.}\
  }\textbf {\bibinfo {volume} {328}},\ \bibinfo {pages} {73} (\bibinfo {year}
  {2008})}\BibitemShut {NoStop}%
\bibitem [{\citenamefont {Gupta}\ \emph {et~al.}(2011)\citenamefont {Gupta},
  \citenamefont {Hampton}, \citenamefont {Stokes}, \citenamefont {Nguyen},\
  and\ \citenamefont {Miller}}]{Gupta_JCIS_Struc_2011}%
  \BibitemOpen
  \bibfield  {author} {\bibinfo {author} {\bibfnamefont {V.}~\bibnamefont
  {Gupta}}, \bibinfo {author} {\bibfnamefont {M.~A.}\ \bibnamefont {Hampton}},
  \bibinfo {author} {\bibfnamefont {J.~R.}\ \bibnamefont {Stokes}}, \bibinfo
  {author} {\bibfnamefont {A.~V.}\ \bibnamefont {Nguyen}},\ and\ \bibinfo
  {author} {\bibfnamefont {J.~D.}\ \bibnamefont {Miller}},\ }\bibfield  {title}
  {\bibinfo {title} {Particle interactions in kaolinite suspensions and
  corresponding aggregate structures},\ }\href
  {https://doi.org/10.1016/j.jcis.2011.03.043} {\bibfield  {journal} {\bibinfo
  {journal} {J. Colloid Interface Sci.}\ }\textbf {\bibinfo {volume} {359}},\
  \bibinfo {pages} {95} (\bibinfo {year} {2011})}\BibitemShut {NoStop}%
\bibitem [{\citenamefont {Lin}\ \emph {et~al.}(2015)\citenamefont {Lin},
  \citenamefont {Phan-Thien}, \citenamefont {Lee},\ and\ \citenamefont
  {Khoo}}]{Phan_Thien_Langmuir_2015}%
  \BibitemOpen
  \bibfield  {author} {\bibinfo {author} {\bibfnamefont {Y.}~\bibnamefont
  {Lin}}, \bibinfo {author} {\bibfnamefont {N.}~\bibnamefont {Phan-Thien}},
  \bibinfo {author} {\bibfnamefont {J.~B.~P.}\ \bibnamefont {Lee}},\ and\
  \bibinfo {author} {\bibfnamefont {B.~C.}\ \bibnamefont {Khoo}},\ }\bibfield
  {title} {\bibinfo {title} {Concentration dependence of yield stress and
  dynamic moduli of kaolinite suspensions},\ }\href
  {https://doi.org/10.1021/acs.langmuir.5b00536} {\bibfield  {journal}
  {\bibinfo  {journal} {Langmuir}\ }\textbf {\bibinfo {volume} {31}},\ \bibinfo
  {pages} {4791} (\bibinfo {year} {2015})}\BibitemShut {NoStop}%
\bibitem [{\citenamefont {Neelakantan}\ \emph {et~al.}(2018)\citenamefont
  {Neelakantan}, \citenamefont {{Vaezi G.}},\ and\ \citenamefont
  {Sanders}}]{Neelakantan_Mineral_2018}%
  \BibitemOpen
  \bibfield  {author} {\bibinfo {author} {\bibfnamefont {R.}~\bibnamefont
  {Neelakantan}}, \bibinfo {author} {\bibfnamefont {F.}~\bibnamefont {{Vaezi
  G.}}},\ and\ \bibinfo {author} {\bibfnamefont {R.~S.}\ \bibnamefont
  {Sanders}},\ }\bibfield  {title} {\bibinfo {title} {Effect of shear on the
  yield stress and aggregate structure of flocculant-dosed, concentrated
  kaolinite suspensions},\ }\href
  {https://doi.org/10.1016/j.mineng.2018.03.016} {\bibfield  {journal}
  {\bibinfo  {journal} {Miner. Eng.}\ }\textbf {\bibinfo {volume} {123}},\
  \bibinfo {pages} {95} (\bibinfo {year} {2018})}\BibitemShut {NoStop}%
\bibitem [{\citenamefont {Mewis}(1979)}]{Mewis_JNNFM_1979}%
  \BibitemOpen
  \bibfield  {author} {\bibinfo {author} {\bibfnamefont {J.}~\bibnamefont
  {Mewis}},\ }\bibfield  {title} {\bibinfo {title} {Thixotropy -- a general
  review},\ }\href {https://doi.org/10.1016/0377-0257(79)87001-9} {\bibfield
  {journal} {\bibinfo  {journal} {J. Nonnewton. Fluid Mech.}\ }\textbf
  {\bibinfo {volume} {6}},\ \bibinfo {pages} {1} (\bibinfo {year}
  {1979})}\BibitemShut {NoStop}%
\bibitem [{\citenamefont {Barnes}(1997)}]{Barnes_JNNFM_1997}%
  \BibitemOpen
  \bibfield  {author} {\bibinfo {author} {\bibfnamefont {H.~A.}\ \bibnamefont
  {Barnes}},\ }\bibfield  {title} {\bibinfo {title} {Thixotropy -- a review},\
  }\href {https://doi.org/10.1016/S0377-0257(97)00004-9} {\bibfield  {journal}
  {\bibinfo  {journal} {J. Nonnewton. Fluid Mech.}\ }\textbf {\bibinfo {volume}
  {70}},\ \bibinfo {pages} {1} (\bibinfo {year} {1997})}\BibitemShut {NoStop}%
\bibitem [{\citenamefont {Mewis}\ and\ \citenamefont
  {Wagner}(2009)}]{Mewis_ACIS_2009}%
  \BibitemOpen
  \bibfield  {author} {\bibinfo {author} {\bibfnamefont {J.}~\bibnamefont
  {Mewis}}\ and\ \bibinfo {author} {\bibfnamefont {N.~J.}\ \bibnamefont
  {Wagner}},\ }\bibfield  {title} {\bibinfo {title} {Thixotropy},\ }\href
  {https://doi.org/10.1016/j.cis.2008.09.005} {\bibfield  {journal} {\bibinfo
  {journal} {Adv. Colloid Interface Sci.}\ }\textbf {\bibinfo {volume}
  {147-148}},\ \bibinfo {pages} {214} (\bibinfo {year} {2009})}\BibitemShut
  {NoStop}%
\bibitem [{\citenamefont {Larson}(2015)}]{Larson_JOR_2015}%
  \BibitemOpen
  \bibfield  {author} {\bibinfo {author} {\bibfnamefont {R.~G.}\ \bibnamefont
  {Larson}},\ }\bibfield  {title} {\bibinfo {title} {Constitutive equations for
  thixotropic fluids},\ }\href {https://doi.org/10.1122/1.4913584} {\bibfield
  {journal} {\bibinfo  {journal} {J. Rheol.}\ }\textbf {\bibinfo {volume}
  {59}},\ \bibinfo {pages} {595} (\bibinfo {year} {2015})}\BibitemShut
  {NoStop}%
\bibitem [{\citenamefont {Larson}\ and\ \citenamefont
  {Wei}(2019)}]{Larson_JOR_2019}%
  \BibitemOpen
  \bibfield  {author} {\bibinfo {author} {\bibfnamefont {R.~G.}\ \bibnamefont
  {Larson}}\ and\ \bibinfo {author} {\bibfnamefont {Y.}~\bibnamefont {Wei}},\
  }\bibfield  {title} {\bibinfo {title} {A review of thixotropy and its
  rheological modeling},\ }\href {https://doi.org/10.1122/1.5055031} {\bibfield
   {journal} {\bibinfo  {journal} {J. Rheol.}\ }\textbf {\bibinfo {volume}
  {63}},\ \bibinfo {pages} {477} (\bibinfo {year} {2019})}\BibitemShut
  {NoStop}%
\bibitem [{\citenamefont {Moore}(1959)}]{Moore_1959}%
  \BibitemOpen
  \bibfield  {author} {\bibinfo {author} {\bibfnamefont {F.}~\bibnamefont
  {Moore}},\ }\bibfield  {title} {\bibinfo {title} {The rheology of ceramic
  slips and bodies},\ }\href
  {https://www.scopus.com/record/display.uri?eid=2-s2.0-0001561959&origin=inward}
  {\bibfield  {journal} {\bibinfo  {journal} {Trans. Br. Ceram. Soc.}\ }\textbf
  {\bibinfo {volume} {58}},\ \bibinfo {pages} {470} (\bibinfo {year}
  {1959})}\BibitemShut {NoStop}%
\bibitem [{\citenamefont {Stickel}\ \emph {et~al.}(2006)\citenamefont
  {Stickel}, \citenamefont {Phillips},\ and\ \citenamefont
  {Powell}}]{Stickel_Model_JOR_2006}%
  \BibitemOpen
  \bibfield  {author} {\bibinfo {author} {\bibfnamefont {J.~J.}\ \bibnamefont
  {Stickel}}, \bibinfo {author} {\bibfnamefont {R.~J.}\ \bibnamefont
  {Phillips}},\ and\ \bibinfo {author} {\bibfnamefont {R.~L.}\ \bibnamefont
  {Powell}},\ }\bibfield  {title} {\bibinfo {title} {A constitutive model for
  microstructure and total stress in particulate suspensions},\ }\href
  {https://doi.org/10.1122/1.2209558} {\bibfield  {journal} {\bibinfo
  {journal} {J. Rheol.}\ }\textbf {\bibinfo {volume} {50}},\ \bibinfo {pages}
  {379} (\bibinfo {year} {2006})}\BibitemShut {NoStop}%
\bibitem [{\citenamefont {Grillet}\ \emph {et~al.}(2009)\citenamefont
  {Grillet}, \citenamefont {Rao}, \citenamefont {Adolf}, \citenamefont
  {Kawaguchi},\ and\ \citenamefont {Mondy}}]{Grillet_JOR_2009}%
  \BibitemOpen
  \bibfield  {author} {\bibinfo {author} {\bibfnamefont {A.~M.}\ \bibnamefont
  {Grillet}}, \bibinfo {author} {\bibfnamefont {R.~R.}\ \bibnamefont {Rao}},
  \bibinfo {author} {\bibfnamefont {D.~B.}\ \bibnamefont {Adolf}}, \bibinfo
  {author} {\bibfnamefont {S.}~\bibnamefont {Kawaguchi}},\ and\ \bibinfo
  {author} {\bibfnamefont {L.~A.}\ \bibnamefont {Mondy}},\ }\bibfield  {title}
  {\bibinfo {title} {Practical application of thixotropic suspension models},\
  }\href {https://doi.org/10.1122/1.3037262} {\bibfield  {journal} {\bibinfo
  {journal} {J. Rheol.}\ }\textbf {\bibinfo {volume} {53}},\ \bibinfo {pages}
  {169} (\bibinfo {year} {2009})}\BibitemShut {NoStop}%
\bibitem [{\citenamefont {Dullaert}\ and\ \citenamefont
  {Mewis}(2006)}]{Mewis_JNNFM_2006}%
  \BibitemOpen
  \bibfield  {author} {\bibinfo {author} {\bibfnamefont {K.}~\bibnamefont
  {Dullaert}}\ and\ \bibinfo {author} {\bibfnamefont {J.}~\bibnamefont
  {Mewis}},\ }\bibfield  {title} {\bibinfo {title} {A structural kinetics model
  for thixotropy},\ }\href {https://doi.org/j.jnnfm.2006.06.002} {\bibfield
  {journal} {\bibinfo  {journal} {J. Nonnewton. Fluid Mech.}\ }\textbf
  {\bibinfo {volume} {139}},\ \bibinfo {pages} {21} (\bibinfo {year}
  {2006})}\BibitemShut {NoStop}%
\bibitem [{\citenamefont {de~Souza~Mendes}(2009)}]{DeSouza_JNNFM_2009}%
  \BibitemOpen
  \bibfield  {author} {\bibinfo {author} {\bibfnamefont {P.~R.}\ \bibnamefont
  {de~Souza~Mendes}},\ }\bibfield  {title} {\bibinfo {title} {Modeling the
  thixotropic behavior of structured fluids},\ }\href
  {https://doi.org/10.1016/j.jnnfm.2009.08.005} {\bibfield  {journal} {\bibinfo
   {journal} {J. Nonnewton. Fluid Mech.}\ }\textbf {\bibinfo {volume} {164}},\
  \bibinfo {pages} {66} (\bibinfo {year} {2009})}\BibitemShut {NoStop}%
\bibitem [{\citenamefont {de~Souza~Mendes}(2011)}]{DeSouza_SoftMatt_2011}%
  \BibitemOpen
  \bibfield  {author} {\bibinfo {author} {\bibfnamefont {P.~R.}\ \bibnamefont
  {de~Souza~Mendes}},\ }\bibfield  {title} {\bibinfo {title} {Thixotropic
  elasto-viscoplastic model for structured fluids},\ }\href
  {https://doi.org/10.1039/C0SM01021A} {\bibfield  {journal} {\bibinfo
  {journal} {Soft Matter}\ }\textbf {\bibinfo {volume} {7}},\ \bibinfo {pages}
  {2471} (\bibinfo {year} {2011})}\BibitemShut {NoStop}%
\bibitem [{\citenamefont {de~Souza~Mendes}\ and\ \citenamefont
  {Thompson}(2012)}]{DeSouza_JNNFM_2012}%
  \BibitemOpen
  \bibfield  {author} {\bibinfo {author} {\bibfnamefont {P.~R.}\ \bibnamefont
  {de~Souza~Mendes}}\ and\ \bibinfo {author} {\bibfnamefont {R.~L.}\
  \bibnamefont {Thompson}},\ }\bibfield  {title} {\bibinfo {title} {A critical
  overview of elasto-viscoplastic thixotropic modeling},\ }\href
  {https://doi.org/10.1016/j.jnnfm.2012.08.006} {\bibfield  {journal} {\bibinfo
   {journal} {J. Nonnewton. Fluid Mech.}\ }\textbf {\bibinfo {volume}
  {187-188}},\ \bibinfo {pages} {8} (\bibinfo {year} {2012})}\BibitemShut
  {NoStop}%
\bibitem [{\citenamefont {de~Souza~Mendes}\ and\ \citenamefont
  {Thompson}(2013)}]{DeSouza_JNNFM_2013}%
  \BibitemOpen
  \bibfield  {author} {\bibinfo {author} {\bibfnamefont {P.~R.}\ \bibnamefont
  {de~Souza~Mendes}}\ and\ \bibinfo {author} {\bibfnamefont {R.~L.}\
  \bibnamefont {Thompson}},\ }\bibfield  {title} {\bibinfo {title} {A unified
  approach to model elasto-viscoplastic thixotropic yield-stress materials and
  apparent yield-stress fluids},\ }\href
  {https://doi.org/10.1007/s00397-013-0699-1} {\bibfield  {journal} {\bibinfo
  {journal} {Rheol. Acta}\ }\textbf {\bibinfo {volume} {52}},\ \bibinfo {pages}
  {673} (\bibinfo {year} {2013})}\BibitemShut {NoStop}%
\bibitem [{\citenamefont {Blackwell}\ and\ \citenamefont
  {Ewoldt}(2014)}]{Blackwell_JNNFM_2014}%
  \BibitemOpen
  \bibfield  {author} {\bibinfo {author} {\bibfnamefont {B.~C.}\ \bibnamefont
  {Blackwell}}\ and\ \bibinfo {author} {\bibfnamefont {R.~H.}\ \bibnamefont
  {Ewoldt}},\ }\bibfield  {title} {\bibinfo {title} {A simple
  thixotropic–viscoelastic constitutive model produces unique signatures in
  large-amplitude oscillatory shear ({LAOS})},\ }\href
  {https://doi.org/https://doi.org/10.1016/j.jnnfm.2014.03.006} {\bibfield
  {journal} {\bibinfo  {journal} {J. Nonnewton. Fluid Mech.}\ }\textbf
  {\bibinfo {volume} {208-209}},\ \bibinfo {pages} {27} (\bibinfo {year}
  {2014})}\BibitemShut {NoStop}%
\bibitem [{\citenamefont {Blackwell}\ and\ \citenamefont
  {Ewoldt}(2016)}]{Blackwell_JNNFM_2016}%
  \BibitemOpen
  \bibfield  {author} {\bibinfo {author} {\bibfnamefont {B.~C.}\ \bibnamefont
  {Blackwell}}\ and\ \bibinfo {author} {\bibfnamefont {R.~H.}\ \bibnamefont
  {Ewoldt}},\ }\bibfield  {title} {\bibinfo {title} {Non-integer asymptotic
  scaling of a thixotropic-viscoelastic model in large-amplitude oscillatory
  shear},\ }\href {https://doi.org/10.1016/j.jnnfm.2015.11.009} {\bibfield
  {journal} {\bibinfo  {journal} {J. Nonnewton. Fluid Mech.}\ }\textbf
  {\bibinfo {volume} {227}},\ \bibinfo {pages} {80} (\bibinfo {year}
  {2016})}\BibitemShut {NoStop}%
\bibitem [{\citenamefont {Armstrong}\ \emph {et~al.}(2016)\citenamefont
  {Armstrong}, \citenamefont {Beris}, \citenamefont {Rogers},\ and\
  \citenamefont {Wagner}}]{Wagner_JOR_2016}%
  \BibitemOpen
  \bibfield  {author} {\bibinfo {author} {\bibfnamefont {M.~J.}\ \bibnamefont
  {Armstrong}}, \bibinfo {author} {\bibfnamefont {A.~N.}\ \bibnamefont
  {Beris}}, \bibinfo {author} {\bibfnamefont {S.~A.}\ \bibnamefont {Rogers}},\
  and\ \bibinfo {author} {\bibfnamefont {N.~J.}\ \bibnamefont {Wagner}},\
  }\bibfield  {title} {\bibinfo {title} {Dynamic shear rheology of a
  thixotropic suspension: Comparison of an improved structure-based model with
  large amplitude oscillatory shear experiments},\ }\href
  {https://doi.org/10.1122/1.4943986} {\bibfield  {journal} {\bibinfo
  {journal} {J. Rheol.}\ }\textbf {\bibinfo {volume} {60}},\ \bibinfo {pages}
  {433} (\bibinfo {year} {2016})}\BibitemShut {NoStop}%
\bibitem [{\citenamefont {Wei}\ \emph {et~al.}(2016)\citenamefont {Wei},
  \citenamefont {Solomon},\ and\ \citenamefont {Larson}}]{Larson_JOR_2016}%
  \BibitemOpen
  \bibfield  {author} {\bibinfo {author} {\bibfnamefont {Y.}~\bibnamefont
  {Wei}}, \bibinfo {author} {\bibfnamefont {M.~J.}\ \bibnamefont {Solomon}},\
  and\ \bibinfo {author} {\bibfnamefont {R.~G.}\ \bibnamefont {Larson}},\
  }\bibfield  {title} {\bibinfo {title} {Quantitative nonlinear thixotropic
  model with stretched exponential response in transient shear flows},\ }\href
  {https://doi.org/10.1122/1.4965228} {\bibfield  {journal} {\bibinfo
  {journal} {J. Rheol.}\ }\textbf {\bibinfo {volume} {60}},\ \bibinfo {pages}
  {1301} (\bibinfo {year} {2016})}\BibitemShut {NoStop}%
\bibitem [{\citenamefont {Wei}\ \emph {et~al.}(2018)\citenamefont {Wei},
  \citenamefont {Solomon},\ and\ \citenamefont {Larson}}]{Larson_JOR_2018}%
  \BibitemOpen
  \bibfield  {author} {\bibinfo {author} {\bibfnamefont {Y.}~\bibnamefont
  {Wei}}, \bibinfo {author} {\bibfnamefont {M.~J.}\ \bibnamefont {Solomon}},\
  and\ \bibinfo {author} {\bibfnamefont {R.~G.}\ \bibnamefont {Larson}},\
  }\bibfield  {title} {\bibinfo {title} {A multimode structural kinetics
  constitutive equation for the transient rheology of thixotropic
  elasto-viscoplastic fluids},\ }\href {https://doi.org/10.1122/1.4996752}
  {\bibfield  {journal} {\bibinfo  {journal} {J. Rheol.}\ }\textbf {\bibinfo
  {volume} {62}},\ \bibinfo {pages} {321} (\bibinfo {year} {2018})}\BibitemShut
  {NoStop}%
\bibitem [{\citenamefont {Agarwal}\ \emph {et~al.}(2021)\citenamefont
  {Agarwal}, \citenamefont {Sharma}, \citenamefont {Shankar},\ and\
  \citenamefont {Joshi}}]{Joshi_JOR_2021}%
  \BibitemOpen
  \bibfield  {author} {\bibinfo {author} {\bibfnamefont {M.}~\bibnamefont
  {Agarwal}}, \bibinfo {author} {\bibfnamefont {S.}~\bibnamefont {Sharma}},
  \bibinfo {author} {\bibfnamefont {V.}~\bibnamefont {Shankar}},\ and\ \bibinfo
  {author} {\bibfnamefont {Y.~M.}\ \bibnamefont {Joshi}},\ }\bibfield  {title}
  {\bibinfo {title} {Distinguishing thixotropy from viscoelasticity},\ }\href
  {https://doi.org/10.1122/8.0000262} {\bibfield  {journal} {\bibinfo
  {journal} {J. Rheol.}\ }\textbf {\bibinfo {volume} {65}},\ \bibinfo {pages}
  {663} (\bibinfo {year} {2021})}\BibitemShut {NoStop}%
\bibitem [{\citenamefont {Joshi}(2022)}]{Joshi_JOR_2022}%
  \BibitemOpen
  \bibfield  {author} {\bibinfo {author} {\bibfnamefont {Y.~M.}\ \bibnamefont
  {Joshi}},\ }\bibfield  {title} {\bibinfo {title} {Thixotropy, nonmonotonic
  stress relaxation, and the second law of thermodynamics},\ }\href
  {https://doi.org/10.1122/8.0000363} {\bibfield  {journal} {\bibinfo
  {journal} {J. Rheol.}\ }\textbf {\bibinfo {volume} {66}},\ \bibinfo {pages}
  {111} (\bibinfo {year} {2022})}\BibitemShut {NoStop}%
\bibitem [{\citenamefont {Sollich}\ \emph {et~al.}(1997)\citenamefont
  {Sollich}, \citenamefont {Lequeux}, \citenamefont {H\'ebraud},\ and\
  \citenamefont {Cates}}]{SGR_PRL_1997}%
  \BibitemOpen
  \bibfield  {author} {\bibinfo {author} {\bibfnamefont {P.}~\bibnamefont
  {Sollich}}, \bibinfo {author} {\bibfnamefont {F.}~\bibnamefont {Lequeux}},
  \bibinfo {author} {\bibfnamefont {P.}~\bibnamefont {H\'ebraud}},\ and\
  \bibinfo {author} {\bibfnamefont {M.~E.}\ \bibnamefont {Cates}},\ }\bibfield
  {title} {\bibinfo {title} {Rheology of soft glassy materials},\ }\href
  {https://doi.org/10.1103/PhysRevLett.78.2020} {\bibfield  {journal} {\bibinfo
   {journal} {Phys. Rev. Lett.}\ }\textbf {\bibinfo {volume} {78}},\ \bibinfo
  {pages} {2020} (\bibinfo {year} {1997})}\BibitemShut {NoStop}%
\bibitem [{\citenamefont {Sollich}(1998)}]{SGR_PRE_1998}%
  \BibitemOpen
  \bibfield  {author} {\bibinfo {author} {\bibfnamefont {P.}~\bibnamefont
  {Sollich}},\ }\bibfield  {title} {\bibinfo {title} {Rheological constitutive
  equation for a model of soft glassy materials},\ }\href
  {https://doi.org/10.1103/PhysRevE.58.738} {\bibfield  {journal} {\bibinfo
  {journal} {Phys. Rev. E}\ }\textbf {\bibinfo {volume} {58}},\ \bibinfo
  {pages} {738} (\bibinfo {year} {1998})}\BibitemShut {NoStop}%
\bibitem [{\citenamefont {Bocquet}\ \emph {et~al.}(2009)\citenamefont
  {Bocquet}, \citenamefont {Colin},\ and\ \citenamefont
  {Ajdari}}]{Bocquet_PRL_2009}%
  \BibitemOpen
  \bibfield  {author} {\bibinfo {author} {\bibfnamefont {L.}~\bibnamefont
  {Bocquet}}, \bibinfo {author} {\bibfnamefont {A.}~\bibnamefont {Colin}},\
  and\ \bibinfo {author} {\bibfnamefont {A.}~\bibnamefont {Ajdari}},\
  }\bibfield  {title} {\bibinfo {title} {Kinetic theory of plastic flow in soft
  glassy materials},\ }\href {https://doi.org/10.1103/PhysRevLett.103.036001}
  {\bibfield  {journal} {\bibinfo  {journal} {Phys. Rev. Lett.}\ }\textbf
  {\bibinfo {volume} {103}},\ \bibinfo {pages} {036001} (\bibinfo {year}
  {2009})}\BibitemShut {NoStop}%
\bibitem [{\citenamefont {Mansard}\ \emph {et~al.}(2011)\citenamefont
  {Mansard}, \citenamefont {Colin}, \citenamefont {Chauduri},\ and\
  \citenamefont {Bocquet}}]{Bocquet_SoftMatt_2011}%
  \BibitemOpen
  \bibfield  {author} {\bibinfo {author} {\bibfnamefont {V.}~\bibnamefont
  {Mansard}}, \bibinfo {author} {\bibfnamefont {A.}~\bibnamefont {Colin}},
  \bibinfo {author} {\bibfnamefont {P.}~\bibnamefont {Chauduri}},\ and\
  \bibinfo {author} {\bibfnamefont {L.}~\bibnamefont {Bocquet}},\ }\bibfield
  {title} {\bibinfo {title} {A kinetic elasto-plastic model exhibiting
  viscosity bifurcation in soft glassy materials},\ }\href
  {https://doi.org/10.1039/C1SM05229B} {\bibfield  {journal} {\bibinfo
  {journal} {Soft Matter}\ }\textbf {\bibinfo {volume} {7}},\ \bibinfo {pages}
  {5524} (\bibinfo {year} {2011})}\BibitemShut {NoStop}%
\bibitem [{\citenamefont {Jeyaseelan}\ and\ \citenamefont
  {Giacomin}(2008)}]{Jeyaseelan_Model_JNNFM_2008}%
  \BibitemOpen
  \bibfield  {author} {\bibinfo {author} {\bibfnamefont {R.~S.}\ \bibnamefont
  {Jeyaseelan}}\ and\ \bibinfo {author} {\bibfnamefont {A.~J.}\ \bibnamefont
  {Giacomin}},\ }\bibfield  {title} {\bibinfo {title} {Network theory for
  polymer solutions in large amplitude oscillatory shear},\ }\href
  {https://doi.org/10.1016/j.jnnfm.2007.04.012} {\bibfield  {journal} {\bibinfo
   {journal} {J. Nonnewton. Fluid Mech.}\ }\textbf {\bibinfo {volume} {148}},\
  \bibinfo {pages} {24} (\bibinfo {year} {2008})}\BibitemShut {NoStop}%
\bibitem [{\citenamefont {Moorcroft}\ \emph {et~al.}(2011)\citenamefont
  {Moorcroft}, \citenamefont {Cates},\ and\ \citenamefont
  {Fielding}}]{Fielding_PRL_2011}%
  \BibitemOpen
  \bibfield  {author} {\bibinfo {author} {\bibfnamefont {R.~L.}\ \bibnamefont
  {Moorcroft}}, \bibinfo {author} {\bibfnamefont {M.~E.}\ \bibnamefont
  {Cates}},\ and\ \bibinfo {author} {\bibfnamefont {S.~M.}\ \bibnamefont
  {Fielding}},\ }\bibfield  {title} {\bibinfo {title} {Age-dependent transient
  shear banding in soft glasses},\ }\href
  {https://doi.org/10.1103/PhysRevLett.106.055502} {\bibfield  {journal}
  {\bibinfo  {journal} {Phys. Rev. Lett.}\ }\textbf {\bibinfo {volume} {106}},\
  \bibinfo {pages} {055502} (\bibinfo {year} {2011})}\BibitemShut {NoStop}%
\bibitem [{\citenamefont {Radhakrishnan}\ \emph {et~al.}(2017)\citenamefont
  {Radhakrishnan}, \citenamefont {Divoux}, \citenamefont {Manneville},\ and\
  \citenamefont {Fielding}}]{Fielding_SM_2017}%
  \BibitemOpen
  \bibfield  {author} {\bibinfo {author} {\bibfnamefont {R.}~\bibnamefont
  {Radhakrishnan}}, \bibinfo {author} {\bibfnamefont {T.}~\bibnamefont
  {Divoux}}, \bibinfo {author} {\bibfnamefont {S.}~\bibnamefont {Manneville}},\
  and\ \bibinfo {author} {\bibfnamefont {S.~M.}\ \bibnamefont {Fielding}},\
  }\bibfield  {title} {\bibinfo {title} {Understanding rheological hysteresis
  in soft glassy materials},\ }\href {https://doi.org/10.1039/C6SM02581A}
  {\bibfield  {journal} {\bibinfo  {journal} {Soft Matter}\ }\textbf {\bibinfo
  {volume} {13}},\ \bibinfo {pages} {1834} (\bibinfo {year}
  {2017})}\BibitemShut {NoStop}%
\bibitem [{\citenamefont {Moorcroft}\ and\ \citenamefont
  {Fielding}(2013)}]{Fielding_PRL_2013}%
  \BibitemOpen
  \bibfield  {author} {\bibinfo {author} {\bibfnamefont {R.~L.}\ \bibnamefont
  {Moorcroft}}\ and\ \bibinfo {author} {\bibfnamefont {S.~M.}\ \bibnamefont
  {Fielding}},\ }\bibfield  {title} {\bibinfo {title} {Criteria for shear
  banding in time-dependent flows of complex fluids},\ }\href
  {https://doi.org/10.1103/PhysRevLett.110.086001} {\bibfield  {journal}
  {\bibinfo  {journal} {Phys. Rev. Lett.}\ }\textbf {\bibinfo {volume} {110}},\
  \bibinfo {pages} {086001} (\bibinfo {year} {2013})}\BibitemShut {NoStop}%
\bibitem [{\citenamefont {Moan}\ \emph {et~al.}(2003)\citenamefont {Moan},
  \citenamefont {Aubry},\ and\ \citenamefont {Bossard}}]{Bossard_JOR_2003}%
  \BibitemOpen
  \bibfield  {author} {\bibinfo {author} {\bibfnamefont {M.}~\bibnamefont
  {Moan}}, \bibinfo {author} {\bibfnamefont {T.}~\bibnamefont {Aubry}},\ and\
  \bibinfo {author} {\bibfnamefont {F.}~\bibnamefont {Bossard}},\ }\bibfield
  {title} {\bibinfo {title} {Nonlinear behavior of very concentrated
  suspensions of plate-like kaolin particles in shear flow},\ }\href
  {https://doi.org/10.1122/1.1608952} {\bibfield  {journal} {\bibinfo
  {journal} {J. Rheol.}\ }\textbf {\bibinfo {volume} {47}},\ \bibinfo {pages}
  {1493} (\bibinfo {year} {2003})}\BibitemShut {NoStop}%
\bibitem [{\citenamefont {Bossard}\ \emph {et~al.}(2007)\citenamefont
  {Bossard}, \citenamefont {Moan},\ and\ \citenamefont
  {Aubry}}]{Bossard_JOR_2007}%
  \BibitemOpen
  \bibfield  {author} {\bibinfo {author} {\bibfnamefont {F.}~\bibnamefont
  {Bossard}}, \bibinfo {author} {\bibfnamefont {M.}~\bibnamefont {Moan}},\ and\
  \bibinfo {author} {\bibfnamefont {T.}~\bibnamefont {Aubry}},\ }\bibfield
  {title} {\bibinfo {title} {Linear and nonlinear viscoelastic behavior of very
  concentrated plate-like kaolin suspensions},\ }\href
  {https://doi.org/10.1122/1.2790023} {\bibfield  {journal} {\bibinfo
  {journal} {J. Rheol.}\ }\textbf {\bibinfo {volume} {51}},\ \bibinfo {pages}
  {1253} (\bibinfo {year} {2007})}\BibitemShut {NoStop}%
\bibitem [{\citenamefont {Joshi}\ \emph {et~al.}(2008)\citenamefont {Joshi},
  \citenamefont {Reddy}, \citenamefont {Kulkarni}, \citenamefont {Kumar},\ and\
  \citenamefont {Chhabra}}]{Joshi_PRSA_2008}%
  \BibitemOpen
  \bibfield  {author} {\bibinfo {author} {\bibfnamefont {Y.~M.}\ \bibnamefont
  {Joshi}}, \bibinfo {author} {\bibfnamefont {G.~K.}\ \bibnamefont {Reddy}},
  \bibinfo {author} {\bibfnamefont {A.~L.}\ \bibnamefont {Kulkarni}}, \bibinfo
  {author} {\bibfnamefont {N.}~\bibnamefont {Kumar}},\ and\ \bibinfo {author}
  {\bibfnamefont {R.~P.}\ \bibnamefont {Chhabra}},\ }\bibfield  {title}
  {\bibinfo {title} {Rheological behaviour of aqueous suspensions of laponite:
  new insights into the ageing phenomena},\ }\href
  {https://doi.org/10.1098/rspa.2007.0250} {\bibfield  {journal} {\bibinfo
  {journal} {Proc. R. Soc. London A}\ }\textbf {\bibinfo {volume} {464}},\
  \bibinfo {pages} {469} (\bibinfo {year} {2008})}\BibitemShut {NoStop}%
\bibitem [{\citenamefont {Joshi}\ and\ \citenamefont
  {Petekidis}(2018)}]{Joshi_RA_2018}%
  \BibitemOpen
  \bibfield  {author} {\bibinfo {author} {\bibfnamefont {Y.~M.}\ \bibnamefont
  {Joshi}}\ and\ \bibinfo {author} {\bibfnamefont {G.}~\bibnamefont
  {Petekidis}},\ }\bibfield  {title} {\bibinfo {title} {Yield stress fluids and
  ageing},\ }\href {https://doi.org/10.1007/s00397-018-1096-6} {\bibfield
  {journal} {\bibinfo  {journal} {Rheol. Acta}\ }\textbf {\bibinfo {volume}
  {57}},\ \bibinfo {pages} {521} (\bibinfo {year} {2018})}\BibitemShut
  {NoStop}%
\bibitem [{\citenamefont {Rueb}\ and\ \citenamefont
  {Zukoski}(1997)}]{Zukoski_JOR_1997}%
  \BibitemOpen
  \bibfield  {author} {\bibinfo {author} {\bibfnamefont {C.~J.}\ \bibnamefont
  {Rueb}}\ and\ \bibinfo {author} {\bibfnamefont {C.~F.}\ \bibnamefont
  {Zukoski}},\ }\bibfield  {title} {\bibinfo {title} {Viscoelastic properties
  of colloidal gels},\ }\href {https://doi.org/10.1122/1.550812} {\bibfield
  {journal} {\bibinfo  {journal} {J. Rheol.}\ }\textbf {\bibinfo {volume}
  {41}},\ \bibinfo {pages} {197} (\bibinfo {year} {1997})}\BibitemShut
  {NoStop}%
\bibitem [{\citenamefont {Abbasi~Moud}\ \emph {et~al.}(2021)\citenamefont
  {Abbasi~Moud}, \citenamefont {Poisson}, \citenamefont {Hudson},\ and\
  \citenamefont {Hatzikiriakos}}]{Kaolin_PoF_2021}%
  \BibitemOpen
  \bibfield  {author} {\bibinfo {author} {\bibfnamefont {A.}~\bibnamefont
  {Abbasi~Moud}}, \bibinfo {author} {\bibfnamefont {J.}~\bibnamefont
  {Poisson}}, \bibinfo {author} {\bibfnamefont {Z.~M.}\ \bibnamefont
  {Hudson}},\ and\ \bibinfo {author} {\bibfnamefont {S.~G.}\ \bibnamefont
  {Hatzikiriakos}},\ }\bibfield  {title} {\bibinfo {title} {Yield stress and
  wall slip of kaolinite networks},\ }\href {https://doi.org/10.1063/5.0050541}
  {\bibfield  {journal} {\bibinfo  {journal} {Phys. Fluids}\ }\textbf {\bibinfo
  {volume} {33}},\ \bibinfo {pages} {053105} (\bibinfo {year}
  {2021})}\BibitemShut {NoStop}%
\bibitem [{\citenamefont {Larson}(1999)}]{Larson_rheology_1999}%
  \BibitemOpen
  \bibfield  {author} {\bibinfo {author} {\bibfnamefont {R.}~\bibnamefont
  {Larson}},\ }\href {https://books.google.com/books?id=BgTjwAEACAAJ} {\emph
  {\bibinfo {title} {The rheology and structure of complex fluids}}}\ (\bibinfo
   {publisher} {Oxford University Press},\ \bibinfo {address} {New York},\
  \bibinfo {year} {1999})\BibitemShut {NoStop}%
\bibitem [{\citenamefont {Augusto}\ \emph {et~al.}(2012)\citenamefont
  {Augusto}, \citenamefont {Falguera}, \citenamefont {Cristianini},\ and\
  \citenamefont {Ibarz}}]{Augusto_Food_2012}%
  \BibitemOpen
  \bibfield  {author} {\bibinfo {author} {\bibfnamefont {P.~E.~D.}\
  \bibnamefont {Augusto}}, \bibinfo {author} {\bibfnamefont {V.}~\bibnamefont
  {Falguera}}, \bibinfo {author} {\bibfnamefont {M.}~\bibnamefont
  {Cristianini}},\ and\ \bibinfo {author} {\bibfnamefont {A.}~\bibnamefont
  {Ibarz}},\ }\bibfield  {title} {\bibinfo {title} {Rheological behavior of
  tomato juice: Steady-state shear and time-dependent modeling},\ }\href
  {https://doi.org/10.1007/s11947-010-0472-8} {\bibfield  {journal} {\bibinfo
  {journal} {Food Bioproc. Technol.}\ }\textbf {\bibinfo {volume} {5}},\
  \bibinfo {pages} {1715} (\bibinfo {year} {2012})}\BibitemShut {NoStop}%
\bibitem [{\citenamefont {Glicerina}\ \emph {et~al.}(2015)\citenamefont
  {Glicerina}, \citenamefont {Balestra}, \citenamefont {Dalla~Rosa},\ and\
  \citenamefont {Romani}}]{Glicerina_Chocolate_2015}%
  \BibitemOpen
  \bibfield  {author} {\bibinfo {author} {\bibfnamefont {V.}~\bibnamefont
  {Glicerina}}, \bibinfo {author} {\bibfnamefont {F.}~\bibnamefont {Balestra}},
  \bibinfo {author} {\bibfnamefont {M.}~\bibnamefont {Dalla~Rosa}},\ and\
  \bibinfo {author} {\bibfnamefont {S.}~\bibnamefont {Romani}},\ }\bibfield
  {title} {\bibinfo {title} {Effect of manufacturing process on the
  microstructural and rheological properties of milk chocolate},\ }\href
  {https://doi.org/10.1016/j.jfoodeng.2014.06.039} {\bibfield  {journal}
  {\bibinfo  {journal} {J. Food Eng.}\ }\textbf {\bibinfo {volume} {145}},\
  \bibinfo {pages} {45} (\bibinfo {year} {2015})}\BibitemShut {NoStop}%
\bibitem [{\citenamefont {Glicerina}\ \emph {et~al.}(2016)\citenamefont
  {Glicerina}, \citenamefont {Balestra}, \citenamefont {Dalla~Rosa},\ and\
  \citenamefont {Romani}}]{Glicerina_Chocolate_2016}%
  \BibitemOpen
  \bibfield  {author} {\bibinfo {author} {\bibfnamefont {V.}~\bibnamefont
  {Glicerina}}, \bibinfo {author} {\bibfnamefont {F.}~\bibnamefont {Balestra}},
  \bibinfo {author} {\bibfnamefont {M.}~\bibnamefont {Dalla~Rosa}},\ and\
  \bibinfo {author} {\bibfnamefont {S.}~\bibnamefont {Romani}},\ }\bibfield
  {title} {\bibinfo {title} {Microstructural and rheological characteristics of
  dark, milk and white chocolate: A comparative study},\ }\href
  {https://doi.org/10.1016/j.jfoodeng.2015.08.011} {\bibfield  {journal}
  {\bibinfo  {journal} {J. Food Eng.}\ }\textbf {\bibinfo {volume} {169}},\
  \bibinfo {pages} {165} (\bibinfo {year} {2016})}\BibitemShut {NoStop}%
\bibitem [{\citenamefont {Krystyjan}\ \emph {et~al.}(2016)\citenamefont
  {Krystyjan}, \citenamefont {Sikora}, \citenamefont {Adamczyk}, \citenamefont
  {Dobosz}, \citenamefont {Tomasik}, \citenamefont {Berski}, \citenamefont
  {Łukasiewicz},\ and\ \citenamefont {Izak}}]{Potato_CP_2016}%
  \BibitemOpen
  \bibfield  {author} {\bibinfo {author} {\bibfnamefont {M.}~\bibnamefont
  {Krystyjan}}, \bibinfo {author} {\bibfnamefont {M.}~\bibnamefont {Sikora}},
  \bibinfo {author} {\bibfnamefont {G.}~\bibnamefont {Adamczyk}}, \bibinfo
  {author} {\bibfnamefont {A.}~\bibnamefont {Dobosz}}, \bibinfo {author}
  {\bibfnamefont {P.}~\bibnamefont {Tomasik}}, \bibinfo {author} {\bibfnamefont
  {W.}~\bibnamefont {Berski}}, \bibinfo {author} {\bibfnamefont
  {M.}~\bibnamefont {Łukasiewicz}},\ and\ \bibinfo {author} {\bibfnamefont
  {P.}~\bibnamefont {Izak}},\ }\bibfield  {title} {\bibinfo {title}
  {Thixotropic properties of waxy potato starch depending on the degree of the
  granules pasting},\ }\href {https://doi.org/10.1016/j.carbpol.2015.12.063}
  {\bibfield  {journal} {\bibinfo  {journal} {Carbohydr. Polym.}\ }\textbf
  {\bibinfo {volume} {141}},\ \bibinfo {pages} {126} (\bibinfo {year}
  {2016})}\BibitemShut {NoStop}%
\bibitem [{\citenamefont {Dimitriou}\ and\ \citenamefont
  {McKinley}(2014)}]{Mckinley_SoftMatt_2014}%
  \BibitemOpen
  \bibfield  {author} {\bibinfo {author} {\bibfnamefont {C.~J.}\ \bibnamefont
  {Dimitriou}}\ and\ \bibinfo {author} {\bibfnamefont {G.~H.}\ \bibnamefont
  {McKinley}},\ }\bibfield  {title} {\bibinfo {title} {A comprehensive
  constitutive law for waxy crude oil: a thixotropic yield stress fluid},\
  }\href {https://doi.org/10.1039/C4SM00578C} {\bibfield  {journal} {\bibinfo
  {journal} {Soft Matter}\ }\textbf {\bibinfo {volume} {10}},\ \bibinfo {pages}
  {6619} (\bibinfo {year} {2014})}\BibitemShut {NoStop}%
\bibitem [{\citenamefont {Geri}\ \emph {et~al.}(2017)\citenamefont {Geri},
  \citenamefont {Venkatesan}, \citenamefont {Sambath},\ and\ \citenamefont
  {McKinley}}]{McKinley_JOR_2017}%
  \BibitemOpen
  \bibfield  {author} {\bibinfo {author} {\bibfnamefont {M.}~\bibnamefont
  {Geri}}, \bibinfo {author} {\bibfnamefont {R.}~\bibnamefont {Venkatesan}},
  \bibinfo {author} {\bibfnamefont {K.}~\bibnamefont {Sambath}},\ and\ \bibinfo
  {author} {\bibfnamefont {G.~H.}\ \bibnamefont {McKinley}},\ }\bibfield
  {title} {\bibinfo {title} {Thermokinematic memory and the thixotropic
  elasto-viscoplasticity of waxy crude oils},\ }\href
  {https://doi.org/10.1122/1.4978259} {\bibfield  {journal} {\bibinfo
  {journal} {J. Rheol.}\ }\textbf {\bibinfo {volume} {61}},\ \bibinfo {pages}
  {427} (\bibinfo {year} {2017})}\BibitemShut {NoStop}%
\bibitem [{\citenamefont {Divoux}\ \emph {et~al.}(2013)\citenamefont {Divoux},
  \citenamefont {Grenard},\ and\ \citenamefont {Manneville}}]{Divoux_PRL_2013}%
  \BibitemOpen
  \bibfield  {author} {\bibinfo {author} {\bibfnamefont {T.}~\bibnamefont
  {Divoux}}, \bibinfo {author} {\bibfnamefont {V.}~\bibnamefont {Grenard}},\
  and\ \bibinfo {author} {\bibfnamefont {S.}~\bibnamefont {Manneville}},\
  }\bibfield  {title} {\bibinfo {title} {Rheological hysteresis in soft glassy
  materials},\ }\href {https://doi.org/10.1103/PhysRevLett.110.018304}
  {\bibfield  {journal} {\bibinfo  {journal} {Phys. Rev. Lett.}\ }\textbf
  {\bibinfo {volume} {110}},\ \bibinfo {pages} {018304} (\bibinfo {year}
  {2013})}\BibitemShut {NoStop}%
\bibitem [{\citenamefont {Puisto}\ \emph {et~al.}(2015)\citenamefont {Puisto},
  \citenamefont {Mohtaschemi}, \citenamefont {Alava},\ and\ \citenamefont
  {Illa}}]{Puisto_PRE_2015}%
  \BibitemOpen
  \bibfield  {author} {\bibinfo {author} {\bibfnamefont {A.}~\bibnamefont
  {Puisto}}, \bibinfo {author} {\bibfnamefont {M.}~\bibnamefont {Mohtaschemi}},
  \bibinfo {author} {\bibfnamefont {M.~J.}\ \bibnamefont {Alava}},\ and\
  \bibinfo {author} {\bibfnamefont {X.}~\bibnamefont {Illa}},\ }\bibfield
  {title} {\bibinfo {title} {Dynamic hysteresis in the rheology of complex
  fluids},\ }\href {https://doi.org/10.1103/PhysRevE.91.042314} {\bibfield
  {journal} {\bibinfo  {journal} {Phys. Rev. E}\ }\textbf {\bibinfo {volume}
  {91}},\ \bibinfo {pages} {042314} (\bibinfo {year} {2015})}\BibitemShut
  {NoStop}%
\bibitem [{\citenamefont {Fielding}(2020)}]{Fielding_JOR_2020}%
  \BibitemOpen
  \bibfield  {author} {\bibinfo {author} {\bibfnamefont {S.~M.}\ \bibnamefont
  {Fielding}},\ }\bibfield  {title} {\bibinfo {title} {Elastoviscoplastic
  rheology and aging in a simplified soft glassy constitutive model},\ }\href
  {https://doi.org/10.1122/1.5140465} {\bibfield  {journal} {\bibinfo
  {journal} {J. Rheol.}\ }\textbf {\bibinfo {volume} {64}},\ \bibinfo {pages}
  {723} (\bibinfo {year} {2020})}\BibitemShut {NoStop}%
\bibitem [{\citenamefont {Jamali}\ \emph {et~al.}(2019)\citenamefont {Jamali},
  \citenamefont {Armstrong},\ and\ \citenamefont
  {McKinley}}]{Mckinley_PRL_2019}%
  \BibitemOpen
  \bibfield  {author} {\bibinfo {author} {\bibfnamefont {S.}~\bibnamefont
  {Jamali}}, \bibinfo {author} {\bibfnamefont {R.~C.}\ \bibnamefont
  {Armstrong}},\ and\ \bibinfo {author} {\bibfnamefont {G.~H.}\ \bibnamefont
  {McKinley}},\ }\bibfield  {title} {\bibinfo {title} {Multiscale nature of
  thixotropy and rheological hysteresis in attractive colloidal suspensions
  under shear},\ }\href {https://doi.org/10.1103/PhysRevLett.123.248003}
  {\bibfield  {journal} {\bibinfo  {journal} {Phys. Rev. Lett.}\ }\textbf
  {\bibinfo {volume} {123}},\ \bibinfo {pages} {248003} (\bibinfo {year}
  {2019})}\BibitemShut {NoStop}%
\bibitem [{\citenamefont {Mujumdar}\ \emph {et~al.}(2002)\citenamefont
  {Mujumdar}, \citenamefont {Beris},\ and\ \citenamefont
  {Metzner}}]{Beris_JNNFM_2002}%
  \BibitemOpen
  \bibfield  {author} {\bibinfo {author} {\bibfnamefont {A.}~\bibnamefont
  {Mujumdar}}, \bibinfo {author} {\bibfnamefont {A.~N.}\ \bibnamefont
  {Beris}},\ and\ \bibinfo {author} {\bibfnamefont {A.~B.}\ \bibnamefont
  {Metzner}},\ }\bibfield  {title} {\bibinfo {title} {Transient phenomena in
  thixotropic systems},\ }\href {https://doi.org/10.1016/S0377-0257(01)00176-8}
  {\bibfield  {journal} {\bibinfo  {journal} {J. Nonnewton. Fluid Mech.}\
  }\textbf {\bibinfo {volume} {102}},\ \bibinfo {pages} {157} (\bibinfo {year}
  {2002})}\BibitemShut {NoStop}%
\bibitem [{\citenamefont {Divoux}\ \emph {et~al.}(2011)\citenamefont {Divoux},
  \citenamefont {Barentin},\ and\ \citenamefont
  {Manneville}}]{Divoux_Soft_Matt_2011}%
  \BibitemOpen
  \bibfield  {author} {\bibinfo {author} {\bibfnamefont {T.}~\bibnamefont
  {Divoux}}, \bibinfo {author} {\bibfnamefont {C.}~\bibnamefont {Barentin}},\
  and\ \bibinfo {author} {\bibfnamefont {S.}~\bibnamefont {Manneville}},\
  }\bibfield  {title} {\bibinfo {title} {From stress-induced fluidization
  processes to herschel-bulkley behaviour in simple yield stress fluids},\
  }\href {https://doi.org/10.1039/C1SM05607G} {\bibfield  {journal} {\bibinfo
  {journal} {Soft Matter}\ }\textbf {\bibinfo {volume} {7}},\ \bibinfo {pages}
  {8409} (\bibinfo {year} {2011})}\BibitemShut {NoStop}%
\bibitem [{\citenamefont {Fourmentin}\ \emph {et~al.}(2015)\citenamefont
  {Fourmentin}, \citenamefont {Ovarlez}, \citenamefont {Faure}, \citenamefont
  {Peter}, \citenamefont {Lesueur}, \citenamefont {Daviller},\ and\
  \citenamefont {Coussot}}]{Fourmentin_Acta_2015}%
  \BibitemOpen
  \bibfield  {author} {\bibinfo {author} {\bibfnamefont {M.}~\bibnamefont
  {Fourmentin}}, \bibinfo {author} {\bibfnamefont {G.}~\bibnamefont {Ovarlez}},
  \bibinfo {author} {\bibfnamefont {P.}~\bibnamefont {Faure}}, \bibinfo
  {author} {\bibfnamefont {U.}~\bibnamefont {Peter}}, \bibinfo {author}
  {\bibfnamefont {D.}~\bibnamefont {Lesueur}}, \bibinfo {author} {\bibfnamefont
  {D.}~\bibnamefont {Daviller}},\ and\ \bibinfo {author} {\bibfnamefont
  {P.}~\bibnamefont {Coussot}},\ }\bibfield  {title} {\bibinfo {title}
  {Rheology of lime paste---a comparison with cement paste},\ }\href
  {https://doi.org/10.1007/s00397-015-0858-7} {\bibfield  {journal} {\bibinfo
  {journal} {Rheol. Acta}\ }\textbf {\bibinfo {volume} {54}},\ \bibinfo {pages}
  {647} (\bibinfo {year} {2015})}\BibitemShut {NoStop}%
\bibitem [{\citenamefont {Perret}\ \emph {et~al.}(1996)\citenamefont {Perret},
  \citenamefont {Locat},\ and\ \citenamefont
  {Martignoni}}]{perret_1996_thixotropic}%
  \BibitemOpen
  \bibfield  {author} {\bibinfo {author} {\bibfnamefont {D.}~\bibnamefont
  {Perret}}, \bibinfo {author} {\bibfnamefont {J.}~\bibnamefont {Locat}},\ and\
  \bibinfo {author} {\bibfnamefont {P.}~\bibnamefont {Martignoni}},\ }\bibfield
   {title} {\bibinfo {title} {Thixotropic behavior during shear of a
  fine-grained mud from eastern canada},\ }\href
  {https://doi.org/10.1016/0013-7952(96)00031-2} {\bibfield  {journal}
  {\bibinfo  {journal} {Eng. Geol.}\ }\textbf {\bibinfo {volume} {43}},\
  \bibinfo {pages} {31} (\bibinfo {year} {1996})}\BibitemShut {NoStop}%
\bibitem [{\citenamefont {Jeong}\ \emph {et~al.}(2015)\citenamefont {Jeong},
  \citenamefont {Locat}, \citenamefont {Torrance},\ and\ \citenamefont
  {Leroueil}}]{jeong_2015_thixotropic}%
  \BibitemOpen
  \bibfield  {author} {\bibinfo {author} {\bibfnamefont {S.~W.}\ \bibnamefont
  {Jeong}}, \bibinfo {author} {\bibfnamefont {J.}~\bibnamefont {Locat}},
  \bibinfo {author} {\bibfnamefont {J.~K.}\ \bibnamefont {Torrance}},\ and\
  \bibinfo {author} {\bibfnamefont {S.}~\bibnamefont {Leroueil}},\ }\bibfield
  {title} {\bibinfo {title} {Thixotropic and anti-thixotropic behaviors of
  fine-grained soils in various flocculated systems},\ }\href
  {https://doi.org/10.1016/j.enggeo.2015.07.014} {\bibfield  {journal}
  {\bibinfo  {journal} {Eng. Geol.}\ }\textbf {\bibinfo {volume} {196}},\
  \bibinfo {pages} {119} (\bibinfo {year} {2015})}\BibitemShut {NoStop}%
\bibitem [{\citenamefont {Labanda}\ and\ \citenamefont
  {Llorens}(2005)}]{labanda_2005_influence}%
  \BibitemOpen
  \bibfield  {author} {\bibinfo {author} {\bibfnamefont {J.}~\bibnamefont
  {Labanda}}\ and\ \bibinfo {author} {\bibfnamefont {J.}~\bibnamefont
  {Llorens}},\ }\bibfield  {title} {\bibinfo {title} {Influence of sodium
  polyacrylate on the rheology of aqueous laponite dispersions},\ }\href
  {https://doi.org/10.1016/j.jcis.2005.03.055} {\bibfield  {journal} {\bibinfo
  {journal} {J. Colloid Interface Sci.}\ }\textbf {\bibinfo {volume} {289}},\
  \bibinfo {pages} {86} (\bibinfo {year} {2005})}\BibitemShut {NoStop}%
\bibitem [{\citenamefont {Jamali}\ and\ \citenamefont
  {McKinley}(2022)}]{Jamali_JoR_2022}%
  \BibitemOpen
  \bibfield  {author} {\bibinfo {author} {\bibfnamefont {S.}~\bibnamefont
  {Jamali}}\ and\ \bibinfo {author} {\bibfnamefont {G.~H.}\ \bibnamefont
  {McKinley}},\ }\bibfield  {title} {\bibinfo {title} {The mnemosyne number and
  the rheology of remembrance},\ }\href {https://doi.org/10.1122/8.0000432}
  {\bibfield  {journal} {\bibinfo  {journal} {J. Rheol.}\ }\textbf {\bibinfo
  {volume} {66}},\ \bibinfo {pages} {1027} (\bibinfo {year}
  {2022})}\BibitemShut {NoStop}%
\bibitem [{\citenamefont {Sharma}\ \emph {et~al.}(2022)\citenamefont {Sharma},
  \citenamefont {Shankar},\ and\ \citenamefont {Joshi}}]{Joshi_ArXiv_2022}%
  \BibitemOpen
  \bibfield  {author} {\bibinfo {author} {\bibfnamefont {S.}~\bibnamefont
  {Sharma}}, \bibinfo {author} {\bibfnamefont {V.}~\bibnamefont {Shankar}},\
  and\ \bibinfo {author} {\bibfnamefont {Y.~M.}\ \bibnamefont {Joshi}},\
  }\bibfield  {title} {\bibinfo {title} {Viscoelasticity and rheological
  hysteresis},\ }\href {https://doi.org/10.48550/arXiv.2203.00890} {\bibfield
  {journal} {\bibinfo  {journal} {arXiv preprint}\ ,\ \bibinfo {pages} {arXiv:
  2203.00890v3}} (\bibinfo {year} {2022})}\BibitemShut {NoStop}%
\bibitem [{\citenamefont {Ewoldt}\ \emph {et~al.}(2008)\citenamefont {Ewoldt},
  \citenamefont {Hosoi},\ and\ \citenamefont {McKinley}}]{Ewoldt_JOR_2008}%
  \BibitemOpen
  \bibfield  {author} {\bibinfo {author} {\bibfnamefont {R.~H.}\ \bibnamefont
  {Ewoldt}}, \bibinfo {author} {\bibfnamefont {A.~E.}\ \bibnamefont {Hosoi}},\
  and\ \bibinfo {author} {\bibfnamefont {G.~H.}\ \bibnamefont {McKinley}},\
  }\bibfield  {title} {\bibinfo {title} {New measures for characterizing
  nonlinear viscoelasticity in large amplitude oscillatory shear},\ }\href
  {https://doi.org/10.1122/1.2970095} {\bibfield  {journal} {\bibinfo
  {journal} {J. Rheol.}\ }\textbf {\bibinfo {volume} {52}},\ \bibinfo {pages}
  {1427} (\bibinfo {year} {2008})}\BibitemShut {NoStop}%
\bibitem [{\citenamefont {Dimitriou}\ \emph {et~al.}(2013)\citenamefont
  {Dimitriou}, \citenamefont {Ewoldt},\ and\ \citenamefont
  {McKinley}}]{Ewoldt_JOR_2013}%
  \BibitemOpen
  \bibfield  {author} {\bibinfo {author} {\bibfnamefont {C.~J.}\ \bibnamefont
  {Dimitriou}}, \bibinfo {author} {\bibfnamefont {R.~H.}\ \bibnamefont
  {Ewoldt}},\ and\ \bibinfo {author} {\bibfnamefont {G.~H.}\ \bibnamefont
  {McKinley}},\ }\bibfield  {title} {\bibinfo {title} {Describing and
  prescribing the constitutive response of yield stress fluids using large
  amplitude oscillatory shear stress ({LAOStress})},\ }\href
  {https://doi.org/10.1122/1.4754023} {\bibfield  {journal} {\bibinfo
  {journal} {J. Rheol.}\ }\textbf {\bibinfo {volume} {57}},\ \bibinfo {pages}
  {27} (\bibinfo {year} {2013})}\BibitemShut {NoStop}%
\bibitem [{\citenamefont {Lee}\ \emph {et~al.}(2019)\citenamefont {Lee},
  \citenamefont {Weigandt}, \citenamefont {Kelley},\ and\ \citenamefont
  {Rogers}}]{SARogers_PRL_2019}%
  \BibitemOpen
  \bibfield  {author} {\bibinfo {author} {\bibfnamefont {J.~C.-W.}\
  \bibnamefont {Lee}}, \bibinfo {author} {\bibfnamefont {K.~M.}\ \bibnamefont
  {Weigandt}}, \bibinfo {author} {\bibfnamefont {E.~G.}\ \bibnamefont
  {Kelley}},\ and\ \bibinfo {author} {\bibfnamefont {S.~A.}\ \bibnamefont
  {Rogers}},\ }\bibfield  {title} {\bibinfo {title} {Structure-property
  relationships via recovery rheology in viscoelastic materials},\ }\href
  {https://doi.org/10.1103/PhysRevLett.122.248003} {\bibfield  {journal}
  {\bibinfo  {journal} {Phys. Rev. Lett.}\ }\textbf {\bibinfo {volume} {122}},\
  \bibinfo {pages} {248003} (\bibinfo {year} {2019})}\BibitemShut {NoStop}%
\bibitem [{\citenamefont {Donley}\ \emph {et~al.}(2020)\citenamefont {Donley},
  \citenamefont {Singh}, \citenamefont {Shetty},\ and\ \citenamefont
  {Rogers}}]{SARogers_PNAS_2020}%
  \BibitemOpen
  \bibfield  {author} {\bibinfo {author} {\bibfnamefont {G.~J.}\ \bibnamefont
  {Donley}}, \bibinfo {author} {\bibfnamefont {P.~K.}\ \bibnamefont {Singh}},
  \bibinfo {author} {\bibfnamefont {A.}~\bibnamefont {Shetty}},\ and\ \bibinfo
  {author} {\bibfnamefont {S.~A.}\ \bibnamefont {Rogers}},\ }\bibfield  {title}
  {\bibinfo {title} {Elucidating the {G"} overshoot in soft materials with a
  yield transition via a time-resolved experimental strain decomposition},\
  }\href {https://doi.org/10.1073/pnas.2003869117} {\bibfield  {journal}
  {\bibinfo  {journal} {Proc. Natl. Acad. Sci. U.S.A.}\ }\textbf {\bibinfo
  {volume} {117}},\ \bibinfo {pages} {21945} (\bibinfo {year}
  {2020})}\BibitemShut {NoStop}%
\bibitem [{\citenamefont {Freund}\ and\ \citenamefont
  {Ewoldt}(2015)}]{Ewoldt_JOR_2015}%
  \BibitemOpen
  \bibfield  {author} {\bibinfo {author} {\bibfnamefont {J.~B.}\ \bibnamefont
  {Freund}}\ and\ \bibinfo {author} {\bibfnamefont {R.~H.}\ \bibnamefont
  {Ewoldt}},\ }\bibfield  {title} {\bibinfo {title} {{Quantitative rheological
  model selection: Good fits versus credible models using Bayesian
  inference}},\ }\href {https://doi.org/10.1122/1.4915299} {\bibfield
  {journal} {\bibinfo  {journal} {J. Rheol.}\ }\textbf {\bibinfo {volume}
  {59}},\ \bibinfo {pages} {667} (\bibinfo {year} {2015})}\BibitemShut
  {NoStop}%
\bibitem [{\citenamefont {Mahmoudabadbozchelou}\ \emph
  {et~al.}(2021)\citenamefont {Mahmoudabadbozchelou}, \citenamefont {Caggioni},
  \citenamefont {Shahsavari}, \citenamefont {Hartt}, \citenamefont
  {Em~Karniadakis},\ and\ \citenamefont {Jamali}}]{Jamali_JOR_2021}%
  \BibitemOpen
  \bibfield  {author} {\bibinfo {author} {\bibfnamefont {M.}~\bibnamefont
  {Mahmoudabadbozchelou}}, \bibinfo {author} {\bibfnamefont {M.}~\bibnamefont
  {Caggioni}}, \bibinfo {author} {\bibfnamefont {S.}~\bibnamefont
  {Shahsavari}}, \bibinfo {author} {\bibfnamefont {W.~H.}\ \bibnamefont
  {Hartt}}, \bibinfo {author} {\bibfnamefont {G.}~\bibnamefont
  {Em~Karniadakis}},\ and\ \bibinfo {author} {\bibfnamefont {S.}~\bibnamefont
  {Jamali}},\ }\bibfield  {title} {\bibinfo {title} {Data-driven
  physics-informed constitutive metamodeling of complex fluids: A multifidelity
  neural network ({MFNN}) framework},\ }\href
  {https://doi.org/10.1122/8.0000138} {\bibfield  {journal} {\bibinfo
  {journal} {J. Rheol.}\ }\textbf {\bibinfo {volume} {65}},\ \bibinfo {pages}
  {179} (\bibinfo {year} {2021})}\BibitemShut {NoStop}%
\bibitem [{\citenamefont {Mahmoudabadbozchelou}\ and\ \citenamefont
  {Jamali}(2021)}]{Jamali_SciRep_2021}%
  \BibitemOpen
  \bibfield  {author} {\bibinfo {author} {\bibfnamefont {M.}~\bibnamefont
  {Mahmoudabadbozchelou}}\ and\ \bibinfo {author} {\bibfnamefont
  {S.}~\bibnamefont {Jamali}},\ }\bibfield  {title} {\bibinfo {title}
  {Rheology-informed neural networks ({RhINNs}) for forward and inverse
  metamodelling of complex fluids},\ }\href
  {https://doi.org/10.1038/s41598-021-91518-3} {\bibfield  {journal} {\bibinfo
  {journal} {Sci. Rep.}\ }\textbf {\bibinfo {volume} {11}},\ \bibinfo {pages}
  {12015} (\bibinfo {year} {2021})}\BibitemShut {NoStop}%
\bibitem [{\citenamefont {Mahmoudabadbozchelou}\ \emph
  {et~al.}(2022{\natexlab{a}})\citenamefont {Mahmoudabadbozchelou},
  \citenamefont {Karniadakis},\ and\ \citenamefont
  {Jamali}}]{Jamali_SoftMatt_2022}%
  \BibitemOpen
  \bibfield  {author} {\bibinfo {author} {\bibfnamefont {M.}~\bibnamefont
  {Mahmoudabadbozchelou}}, \bibinfo {author} {\bibfnamefont {G.~E.}\
  \bibnamefont {Karniadakis}},\ and\ \bibinfo {author} {\bibfnamefont
  {S.}~\bibnamefont {Jamali}},\ }\bibfield  {title} {\bibinfo {title}
  {{nn-PINNs}: Non-newtonian physics-informed neural networks for complex fluid
  modeling},\ }\href {https://doi.org/10.1039/D1SM01298C} {\bibfield  {journal}
  {\bibinfo  {journal} {Soft Matter}\ }\textbf {\bibinfo {volume} {18}},\
  \bibinfo {pages} {172} (\bibinfo {year} {2022}{\natexlab{a}})}\BibitemShut
  {NoStop}%
\bibitem [{\citenamefont {Mahmoudabadbozchelou}\ \emph
  {et~al.}(2022{\natexlab{b}})\citenamefont {Mahmoudabadbozchelou},
  \citenamefont {Kamani}, \citenamefont {Rogers},\ and\ \citenamefont
  {Jamali}}]{Jamali_PNAS_2022}%
  \BibitemOpen
  \bibfield  {author} {\bibinfo {author} {\bibfnamefont {M.}~\bibnamefont
  {Mahmoudabadbozchelou}}, \bibinfo {author} {\bibfnamefont {K.~M.}\
  \bibnamefont {Kamani}}, \bibinfo {author} {\bibfnamefont {S.~A.}\
  \bibnamefont {Rogers}},\ and\ \bibinfo {author} {\bibfnamefont
  {S.}~\bibnamefont {Jamali}},\ }\bibfield  {title} {\bibinfo {title} {Digital
  rheometer twins: Learning the hidden rheology of complex fluids through
  rheology-informed graph neural networks},\ }\href
  {https://doi.org/10.1073/pnas.2202234119} {\bibfield  {journal} {\bibinfo
  {journal} {Proc. Natl. Acad. Sci.}\ }\textbf {\bibinfo {volume} {119}},\
  \bibinfo {pages} {e2202234119} (\bibinfo {year}
  {2022}{\natexlab{b}})}\BibitemShut {NoStop}%
\bibitem [{\citenamefont {Seiphoori}\ \emph {et~al.}(2021)\citenamefont
  {Seiphoori}, \citenamefont {Gunn}, \citenamefont {Kosgodagan~Acharige},
  \citenamefont {Arratia},\ and\ \citenamefont
  {Jerolmack}}]{Seiphoori_GRL_2021}%
  \BibitemOpen
  \bibfield  {author} {\bibinfo {author} {\bibfnamefont {A.}~\bibnamefont
  {Seiphoori}}, \bibinfo {author} {\bibfnamefont {A.}~\bibnamefont {Gunn}},
  \bibinfo {author} {\bibfnamefont {S.}~\bibnamefont {Kosgodagan~Acharige}},
  \bibinfo {author} {\bibfnamefont {P.~E.}\ \bibnamefont {Arratia}},\ and\
  \bibinfo {author} {\bibfnamefont {D.~J.}\ \bibnamefont {Jerolmack}},\
  }\bibfield  {title} {\bibinfo {title} {Tuning sedimentation through surface
  charge and particle shape},\ }\href {https://doi.org/10.1029/2020GL091251}
  {\bibfield  {journal} {\bibinfo  {journal} {Geophys. Res. Lett.}\ }\textbf
  {\bibinfo {volume} {48}},\ \bibinfo {pages} {e2020GL091251} (\bibinfo {year}
  {2021})}\BibitemShut {NoStop}%
\bibitem [{\citenamefont {Marjoram}\ \emph {et~al.}(2003)\citenamefont
  {Marjoram}, \citenamefont {Molitor}, \citenamefont {Plagnol},\ and\
  \citenamefont {Tavar{\'e}}}]{MCMC_PNAS_2003}%
  \BibitemOpen
  \bibfield  {author} {\bibinfo {author} {\bibfnamefont {P.}~\bibnamefont
  {Marjoram}}, \bibinfo {author} {\bibfnamefont {J.}~\bibnamefont {Molitor}},
  \bibinfo {author} {\bibfnamefont {V.}~\bibnamefont {Plagnol}},\ and\ \bibinfo
  {author} {\bibfnamefont {S.}~\bibnamefont {Tavar{\'e}}},\ }\bibfield  {title}
  {\bibinfo {title} {{Markov chain Monte Carlo without likelihoods}},\ }\href
  {https://doi.org/10.1073/pnas.0306899100} {\bibfield  {journal} {\bibinfo
  {journal} {Proc. Natl. Acad. Sci. U.S.A.}\ }\textbf {\bibinfo {volume}
  {100}},\ \bibinfo {pages} {15324} (\bibinfo {year} {2003})}\BibitemShut
  {NoStop}%
\bibitem [{\citenamefont {Metropolis}\ \emph {et~al.}(1953)\citenamefont
  {Metropolis}, \citenamefont {Rosenbluth}, \citenamefont {Rosenbluth},
  \citenamefont {Teller},\ and\ \citenamefont {Teller}}]{Metropolis_1953}%
  \BibitemOpen
  \bibfield  {author} {\bibinfo {author} {\bibfnamefont {N.}~\bibnamefont
  {Metropolis}}, \bibinfo {author} {\bibfnamefont {A.~W.}\ \bibnamefont
  {Rosenbluth}}, \bibinfo {author} {\bibfnamefont {M.~N.}\ \bibnamefont
  {Rosenbluth}}, \bibinfo {author} {\bibfnamefont {A.~H.}\ \bibnamefont
  {Teller}},\ and\ \bibinfo {author} {\bibfnamefont {E.}~\bibnamefont
  {Teller}},\ }\bibfield  {title} {\bibinfo {title} {Equation of state
  calculations by fast computing machines},\ }\href
  {https://doi.org/10.1063/1.1699114} {\bibfield  {journal} {\bibinfo
  {journal} {J. Chem. Phys.}\ }\textbf {\bibinfo {volume} {21}},\ \bibinfo
  {pages} {1087} (\bibinfo {year} {1953})}\BibitemShut {NoStop}%
\bibitem [{\citenamefont {Hastings}(1970)}]{Hastings_1970}%
  \BibitemOpen
  \bibfield  {author} {\bibinfo {author} {\bibfnamefont {W.~K.}\ \bibnamefont
  {Hastings}},\ }\bibfield  {title} {\bibinfo {title} {{Monte Carlo sampling
  methods using Markov chains and their applications}},\ }\href
  {https://doi.org/10.1093/biomet/57.1.97} {\bibfield  {journal} {\bibinfo
  {journal} {Biometrika}\ }\textbf {\bibinfo {volume} {57}},\ \bibinfo {pages}
  {97} (\bibinfo {year} {1970})}\BibitemShut {NoStop}%
\bibitem [{\citenamefont {Haario}\ \emph {et~al.}(2001)\citenamefont {Haario},
  \citenamefont {Saksman},\ and\ \citenamefont
  {Tamminen}}]{Adaptive_MCMC_2001}%
  \BibitemOpen
  \bibfield  {author} {\bibinfo {author} {\bibfnamefont {H.}~\bibnamefont
  {Haario}}, \bibinfo {author} {\bibfnamefont {E.}~\bibnamefont {Saksman}},\
  and\ \bibinfo {author} {\bibfnamefont {J.}~\bibnamefont {Tamminen}},\
  }\bibfield  {title} {\bibinfo {title} {An adaptive {Metropolis} algorithm},\
  }\href {https://doi.org/10.2307/3318737} {\bibfield  {journal} {\bibinfo
  {journal} {Bernoulli}\ }\textbf {\bibinfo {volume} {7}},\ \bibinfo {pages}
  {223} (\bibinfo {year} {2001})}\BibitemShut {NoStop}%
\bibitem [{\citenamefont {Andrieu}\ and\ \citenamefont
  {Thoms}(2008)}]{Adaptive_MCMC_2008}%
  \BibitemOpen
  \bibfield  {author} {\bibinfo {author} {\bibfnamefont {C.}~\bibnamefont
  {Andrieu}}\ and\ \bibinfo {author} {\bibfnamefont {J.}~\bibnamefont
  {Thoms}},\ }\bibfield  {title} {\bibinfo {title} {A tutorial on adaptive
  {MCMC}},\ }\href {https://doi.org/10.1007/s11222-008-9110-y} {\bibfield
  {journal} {\bibinfo  {journal} {Stat. Comput.}\ }\textbf {\bibinfo {volume}
  {18}},\ \bibinfo {pages} {343} (\bibinfo {year} {2008})}\BibitemShut
  {NoStop}%
\bibitem [{\citenamefont {Oliver}\ \emph {et~al.}(2015)\citenamefont {Oliver},
  \citenamefont {Terejanu}, \citenamefont {Simmons},\ and\ \citenamefont
  {Moser}}]{Unobserved_2015}%
  \BibitemOpen
  \bibfield  {author} {\bibinfo {author} {\bibfnamefont {T.~A.}\ \bibnamefont
  {Oliver}}, \bibinfo {author} {\bibfnamefont {G.}~\bibnamefont {Terejanu}},
  \bibinfo {author} {\bibfnamefont {C.~S.}\ \bibnamefont {Simmons}},\ and\
  \bibinfo {author} {\bibfnamefont {R.~D.}\ \bibnamefont {Moser}},\ }\bibfield
  {title} {\bibinfo {title} {Validating predictions of unobserved quantities},\
  }\href {https://doi.org/10.1016/j.cma.2014.08.023} {\bibfield  {journal}
  {\bibinfo  {journal} {Comput. Methods Appl. Mech. Eng.}\ }\textbf {\bibinfo
  {volume} {283}},\ \bibinfo {pages} {1310} (\bibinfo {year}
  {2015})}\BibitemShut {NoStop}%
\bibitem [{\citenamefont {Herschel}\ and\ \citenamefont
  {Bulkley}(1926)}]{Herschel_Bulkley_1926}%
  \BibitemOpen
  \bibfield  {author} {\bibinfo {author} {\bibfnamefont {W.~H.}\ \bibnamefont
  {Herschel}}\ and\ \bibinfo {author} {\bibfnamefont {R.}~\bibnamefont
  {Bulkley}},\ }\bibfield  {title} {\bibinfo {title} {{Konsistenzmessungen von
  Gummi-Benzoll{\"o}sungen}},\ }\href {https://doi.org/10.1007/BF01432034}
  {\bibfield  {journal} {\bibinfo  {journal} {Kolloid-Z.}\ }\textbf {\bibinfo
  {volume} {39}},\ \bibinfo {pages} {291} (\bibinfo {year} {1926})}\BibitemShut
  {NoStop}%
\bibitem [{\citenamefont {Joshi}(2015)}]{Joshi_SM_2015}%
  \BibitemOpen
  \bibfield  {author} {\bibinfo {author} {\bibfnamefont {Y.~M.}\ \bibnamefont
  {Joshi}},\ }\bibfield  {title} {\bibinfo {title} {A model for aging under
  deformation field{,} residual stresses and strains in soft glassy
  materials},\ }\href {http://dx.doi.org/10.1039/C5SM00217F} {\bibfield
  {journal} {\bibinfo  {journal} {Soft Matter}\ }\textbf {\bibinfo {volume}
  {11}},\ \bibinfo {pages} {3198} (\bibinfo {year} {2015})}\BibitemShut
  {NoStop}%
\bibitem [{\citenamefont {Rand}\ and\ \citenamefont
  {Melton}(1977)}]{Rand_JCIS_1977}%
  \BibitemOpen
  \bibfield  {author} {\bibinfo {author} {\bibfnamefont {B.}~\bibnamefont
  {Rand}}\ and\ \bibinfo {author} {\bibfnamefont {I.~E.}\ \bibnamefont
  {Melton}},\ }\bibfield  {title} {\bibinfo {title} {Particle interactions in
  aqueous kaolinite suspensions: I. effect of ph and electrolyte upon the mode
  of particle interaction in homoionic sodium kaolinite suspensions},\ }\href
  {https://doi.org/https://doi.org/10.1016/0021-9797(77)90290-9} {\bibfield
  {journal} {\bibinfo  {journal} {J. Colloid Interface Sci.}\ }\textbf
  {\bibinfo {volume} {60}},\ \bibinfo {pages} {308} (\bibinfo {year}
  {1977})}\BibitemShut {NoStop}%
\bibitem [{\citenamefont {Chattopadhyay}\ \emph {et~al.}(2022)\citenamefont
  {Chattopadhyay}, \citenamefont {Nagaraja},\ and\ \citenamefont
  {Majumdar}}]{chattopadhyay2022effect}%
  \BibitemOpen
  \bibfield  {author} {\bibinfo {author} {\bibfnamefont {S.}~\bibnamefont
  {Chattopadhyay}}, \bibinfo {author} {\bibfnamefont {S.}~\bibnamefont
  {Nagaraja}},\ and\ \bibinfo {author} {\bibfnamefont {S.}~\bibnamefont
  {Majumdar}},\ }\bibfield  {title} {\bibinfo {title} {Effect of adhesive
  interaction on strain stiffening and dissipation in granular gels undergoing
  yielding},\ }\href {https://doi.org/10.1038/s42005-022-00904-4} {\bibfield
  {journal} {\bibinfo  {journal} {Commun. Phys.}\ }\textbf {\bibinfo {volume}
  {5}},\ \bibinfo {pages} {126} (\bibinfo {year} {2022})}\BibitemShut {NoStop}%
\bibitem [{\citenamefont {Dullaert}\ and\ \citenamefont
  {Mewis}(2005)}]{Mewis_JOR_2005}%
  \BibitemOpen
  \bibfield  {author} {\bibinfo {author} {\bibfnamefont {K.}~\bibnamefont
  {Dullaert}}\ and\ \bibinfo {author} {\bibfnamefont {J.}~\bibnamefont
  {Mewis}},\ }\bibfield  {title} {\bibinfo {title} {Thixotropy: Build-up and
  breakdown curves during flow},\ }\href {https://doi.org/10.1122/1.2039868}
  {\bibfield  {journal} {\bibinfo  {journal} {J. Rheol.}\ }\textbf {\bibinfo
  {volume} {49}},\ \bibinfo {pages} {1213} (\bibinfo {year}
  {2005})}\BibitemShut {NoStop}%
\bibitem [{\citenamefont {Wei}\ \emph {et~al.}(2019)\citenamefont {Wei},
  \citenamefont {Solomon},\ and\ \citenamefont
  {Larson}}]{Larson_SoftMatt_2019}%
  \BibitemOpen
  \bibfield  {author} {\bibinfo {author} {\bibfnamefont {Y.}~\bibnamefont
  {Wei}}, \bibinfo {author} {\bibfnamefont {M.~J.}\ \bibnamefont {Solomon}},\
  and\ \bibinfo {author} {\bibfnamefont {R.~G.}\ \bibnamefont {Larson}},\
  }\bibfield  {title} {\bibinfo {title} {Time-dependent shear rate
  inhomogeneities and shear bands in a thixotropic yield-stress fluid under
  transient shear},\ }\href {https://doi.org/10.1039/C9SM00902G} {\bibfield
  {journal} {\bibinfo  {journal} {Soft Matter}\ }\textbf {\bibinfo {volume}
  {15}},\ \bibinfo {pages} {7956} (\bibinfo {year} {2019})}\BibitemShut
  {NoStop}%
\bibitem [{\citenamefont {Hunter}\ and\ \citenamefont
  {Weeks}(2012)}]{Weeks_RPP_2012}%
  \BibitemOpen
  \bibfield  {author} {\bibinfo {author} {\bibfnamefont {G.~L.}\ \bibnamefont
  {Hunter}}\ and\ \bibinfo {author} {\bibfnamefont {E.~R.}\ \bibnamefont
  {Weeks}},\ }\bibfield  {title} {\bibinfo {title} {The physics of the
  colloidal glass transition},\ }\href
  {https://doi.org/10.1088/0034-4885/75/6/066501} {\bibfield  {journal}
  {\bibinfo  {journal} {Rep. Prog. Phys.}\ }\textbf {\bibinfo {volume} {75}},\
  \bibinfo {pages} {066501} (\bibinfo {year} {2012})}\BibitemShut {NoStop}%
\bibitem [{\citenamefont {Pradeep}\ and\ \citenamefont
  {Hsiao}(2020)}]{Pradeep_SM_2020}%
  \BibitemOpen
  \bibfield  {author} {\bibinfo {author} {\bibfnamefont {S.}~\bibnamefont
  {Pradeep}}\ and\ \bibinfo {author} {\bibfnamefont {L.~C.}\ \bibnamefont
  {Hsiao}},\ }\bibfield  {title} {\bibinfo {title} {Contact criterion for
  suspensions of smooth and rough colloids},\ }\href
  {https://doi.org/10.1039/D0SM00072H} {\bibfield  {journal} {\bibinfo
  {journal} {Soft Matter}\ }\textbf {\bibinfo {volume} {16}},\ \bibinfo {pages}
  {4980} (\bibinfo {year} {2020})}\BibitemShut {NoStop}%
\end{thebibliography}%

\end{document}